# Arc Plasma Torch Modeling

*J. P. Trelles, C. Chazelas, A. Vardelle and J. V. R. Heberlein*

**Abstract**

*Arc plasma torches are the primary components of various industrial thermal plasma processes involving plasma spraying, metal cutting and welding, thermal plasma CVD, metal melting and remelting, waste treatment and gas production. They are relatively simple devices whose operation implies intricate thermal, chemical, electrical, and fluid dynamics phenomena. Modeling may be used as a means to better understand the physical processes involved in their operation. This paper presents an overview of the main aspects involved in the modeling of DC arc plasma torches: the mathematical models including thermodynamic and chemical non-equilibrium models, turbulent and radiative transport, thermodynamic and transport property calculation, boundary conditions and arc reattachment models. It focuses on the conventional plasma torches used for plasma spraying that include a hot –cathode and a nozzle anode.*

## Nomenclature

| | | | |
|---|---|---|---|
| **A** | magnetic vector potential | $k_\lambda$ | spectral absorption coefficient |
| $a_{s,r}$ | stoichiometric coefficient in the forward direction for the reaction *r* | $k_{f,r}$ | forward reaction rate for reaction *r* |
| **B** | magnetic field | $k_{b,r}$ | backward reaction rate for reaction *r* |
| $b_{s,r}$ | stoichiometric coefficients in the backward direction for the reaction *r* | $M_s$ | molecular weight of species *s* |
| $C_\mu$ | constant of $k$-$\varepsilon$ model | $m_e$ | electron mass |
| $C_{\varepsilon 1}$ | constant of $k$-$\varepsilon$ model | $n_e$ | Electron number density |
| $C_{\varepsilon 2}$ | constant of $k$-$\varepsilon$ model | $ns$ | number of species |
| $D_s$ | effective diffusivity of species *s* | $nr$ | number of reactions |

| | |
|---|---|
| $D_{sj}$ | binary diffusion coefficient species $s$ and $j$ |
| $\mathbf{E}$ | real electric field |
| $E_b$ | critical electric field |
| $\mathbf{E_p}$ | effective electric field |
| $e$ | elementary electric charge |
| $G_k$ | generation of turbulent kinetic energy |
| $h$ | plasma specific enthalpy |
| $h_e$ | electron specific enthalpy |
| $h_h$ | heavy particles specific enthalpy |
| $h_w$ | heat transfer coefficient |
| $I$ | Arc current intensity |
| $I_\lambda$ | spectral intensity |
| $I_{b\lambda}$ | spectral black body intensity |
| $J_{cath}$ | current density over the cathode surface |
| $J_q$ | arc current density |
| $\mathbf{J}_s$ | mass diffusion flux of species s |
| $k$ | turbulent kinetic energy |
| $k$ | total thermal conductivity |
| $k_B$ | Boltzmann constant |
| $k_e$ | electron translational thermal conductivity |
| $k_h$ | heavy particles thermal conductivity |
| $k_r$ | reactive thermal conductivity |
| $p$ | pressure |
| $\dot{Q}_{eh}$ | electron – heavy-particle energy exchange |
| $\dot{Q}_r$ | volumetric net radiation losses |
| $q_a$ | heat transferred to the anode surface |
| $q_r$ | radiative heat flux |
| $q_{wall}$ | heat transferred to the wall |
| $q'$ | total heat flux |
| $r$ | radius |
| $R_s$ | gas constant of species s |
| $T$ | temperature |
| $T_w$ | wall temperature |
| $T_c$ | critical temperature |
| $T_e$ | electron temperature |
| $T_h$ | heavy particle temperature |
| $t$ | time |
| $\mathbf{u}$ | mass average velocity |
| $\mathbf{u_s}$ | velocity of species s |
| $W_a$ | work function of the anode material |
| $\mathbf{x}$ | spatial coordinate |

### Greek Letters

| | |
|---|---|
| $\delta$ | Kronecker delta |
| $\delta_{es}$ | inelastic collision factor |
| $\varepsilon$ | turbulent dissipation rate |
| $\varepsilon_r$ | effective net emission coefficient |
| $\theta$ | departure from thermal equilibrium |
| $\phi_p$ | effective electric potential |

### Subscripts

| | |
|---|---|
| $a$ | anode |
| $e$ | electron |
| $h$ | heavy particle |
| $j$ | specie |
| $k$ | turbulent kinetic energy |
| $r$ | reactants or reactions |

| | | | |
|---|---|---|---|
| $\lambda$ | wavelength | $s$ | Species |
| $\mu$ | molecular dynamic viscosity | $t$ | turbulent |
| $\mu_0$ | permeability of free space | $w$ | wall |
| $\mu_t$ | turbulent dynamic viscosity | $\varepsilon$ | turbulent dissipation rate |
| $\rho$ | mass density | | |
| $\rho_s$ | mass density of species $s$ | | |
| $\dot{\rho}_s^c$ | volumetric production rate of species $s$ | | |
| $\sigma$ | Electrical conductivity | | |
| $\sigma_k$ | constant of $k$-$\varepsilon$ model | | |
| $\boldsymbol{\tau}$ | stress tensor | | |
| $\nu_{es}$ | collision frequency between electrons and species $s$ | | |
| $\omega$ | reaction molar rate | | |
| $\psi$ | conserved property | | |
| $\dot{\varpi}_r$ | progress rate of reaction $r$ | | |
| $\Delta$ | Finite change in quantity | | |

## Acronyms

| | |
|---|---|
| CFD | Computational Fluid Dynamics |
| DC | Direct Current |
| DES | Detached Eddy Simulation |
| SCEBD | Self-Consistent Effective Binary Diffusion |
| DNS | Direct Numerical Simulation |
| DOM | Discrete Ordinates Methods |
| RANS | Reynolds-Averaged Navier-Stokes |
| EEDF | Electron Energy Distribution Function |
| LES | Large Eddy Simulations |
| LTE | Local Thermal Equilibrium |
| NEC | Net Emission Coefficient |
| NLTE | Non Local Thermal Equilibrium |
| RTE | Radiative Transfer Equation |
| HVOF | High Velocity Oxygen Fuel |

# 1. Introduction

Thermal plasma processes have proven their technological advantage in a wide variety of fields for over 40 years. The features that make thermal plasmas attractive are a high energy density ( ~ $10^6$-$10^7$ J/m$^3$) that comes with high heat flux density (~ $10^7$-$10^9$W/m$^2$), high quenching rate (~ $10^6$-$10^8$ K/s), and high processing rates.

Direct current (DC) arc plasma torches are, generally, the primary component of these processes that include plasma spraying, ultra fine particle synthesis, metal welding and cutting but also, extractive metallurgy, waste treatment and biogas production. These torches operate as thermal, chemical, and electrical devices in processes that achieve material modifications which often cannot be achieved, or are not economically feasible, with other devices. A distinctive example of an application that relies on DC arc plasma torches is plasma spraying that has become a well-established and widely used technology to manufacture coatings resistant to wear, corrosion and temperature and generate near-net shapes of metallic and ceramic parts. For instance, plasma-sprayed coatings make possible turbine blades to withstand temperatures up to 1200 °C and provide unparalleled wear-resistance to prosthetic implants. The continuous development of thermal plasma-based technologies stresses the need for a better understanding of the operation of arc plasma torches.

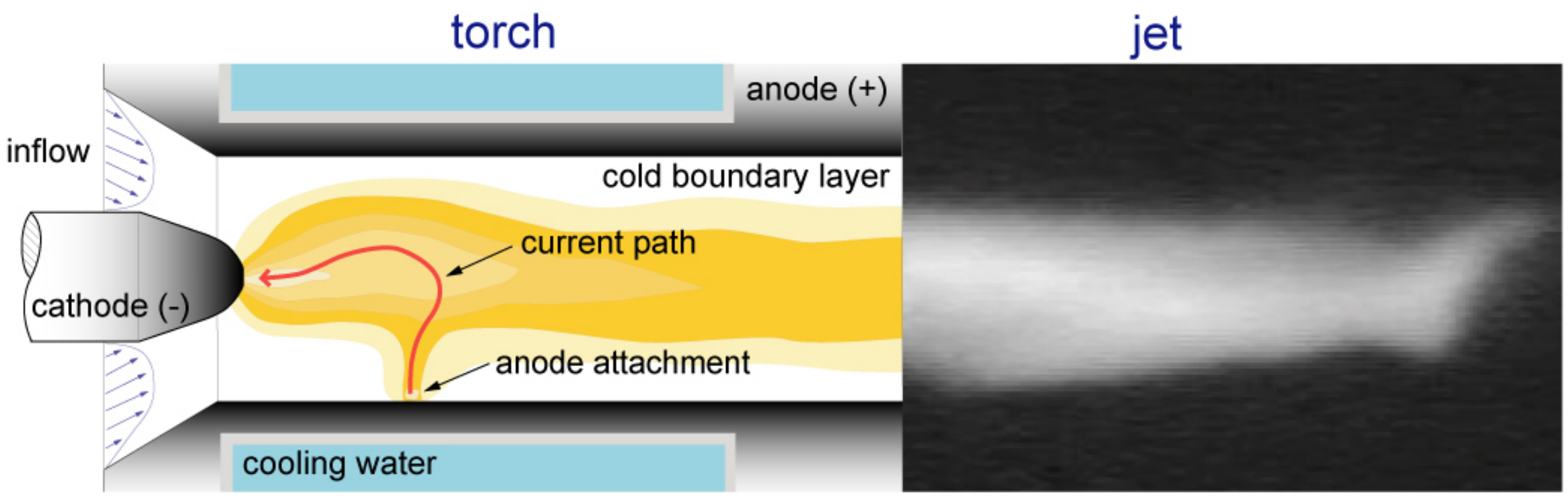

**Fig. 1** Scheme of the plasma flow inside a non-transferred DC arc plasma torch and high speed image of the plasma jet; the diameter of the anode at the torch exit is typically ~ 6 – 10 mm

The apparent simplicity of a DC arc plasma torch is in marked contrast with the complexity of the electrical, chemical, and thermal phenomena involved in its operation. Most DC arc torches have three main components: The cathode, the plasma-forming gas injection stage, and the anode. The anode usually also acts as arc constrictor in so-called non-transferred arc torches (Fig. 1) or forms part of the processing material outside the torch in transferred arc torches (Fig. 2) (Ref 1). Non-transferred arc torches are typically used in applications that rely on the formation of a plasma jet with moderate to very high velocity and, its use as a heat source, high–temperature processing medium or source of specific reactive species, such as plasma spraying and powder synthesis. Transferred arc torches are mostly used in applications that maximize the utilization of heat from the plasma, such as plasma cutting, welding and extractive metallurgy.

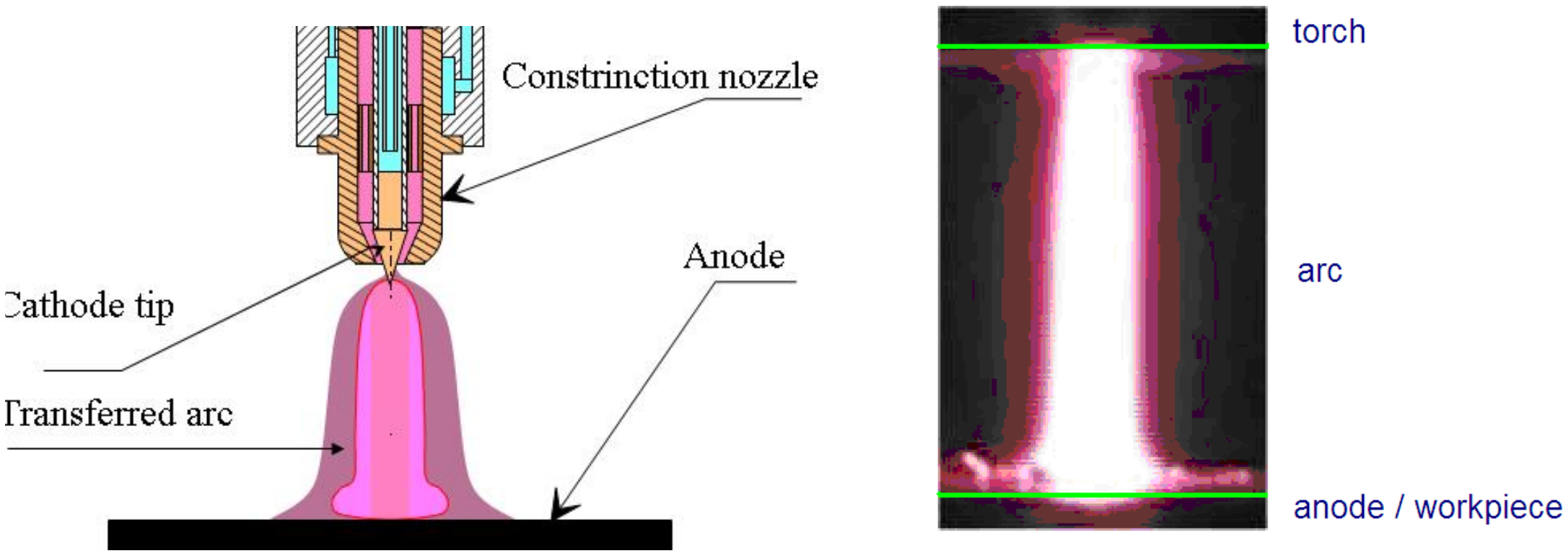

**Fig. 2** Electric arc formed by a transferred arc plasma torch and typical picture

Inside the torch, the working gas flows around the cathode and through a constricting tube or nozzle. The plasma is usually initiated by a high voltage pulse which creates a conductive path for an electric arc to form between the cathode and anode (the torch nozzle in non-transferred arc torches or the working piece in transferred ones). The electric heating produced by the arc causes the gas to reach very high temperatures (e.g. > 10000 K), thus to dissociate and ionize. The cold gas around the surface of the water cooled nozzle or constrictor tube, being electrically non-conductive, constricts the plasma, raising its temperature and velocity. Most of the commercial plasma spray torches operate at atmospheric pressure with electric power levels ranging between 10 and 100 kW, arc currents between 250 and 1000 A, arc voltages between 30 and 100 V, and flow rates between 20 and 150 slpm (standard liters per minute). Common gases used in thermal plasma processing are Ar, He, $H_2$, $N_2$, $O_2$, and mixtures of these.

The modeling of DC arc plasma torches is extremely challenging because the plasma flow is highly nonlinear, presents strong property gradients, is characterized by a wide range of time and length scales, and often includes chemical and thermodynamic non-equilibrium effects, especially near its boundaries. Moreover, the modeling of the torch is frequently part of the modeling of a given plasma process or application (e.g. plasma spraying, plasma cutting). In that case, the description of the plasma flow needs to be coupled with suitable models of the processing material (e.g. sprayed powder, work piece) and its interaction with the plasma flow. Hence, the domain of analysis is typically extended beyond the plasma and across the torch components (e.g. to model the electrical characteristics across the electrodes, or the dissipation of heat to the cooling water) and/or the working material (e.g. to describe heat transferred and phase changes in the processing powder or work piece).

This paper presents an overview of the main aspects involved in DC arc plasma torch modeling, as well as some examples that typify the current state of the art. Particular emphasis is given to conventional non-transferred arc plasma torches with thermionic (hot) cathodes as those employed in plasma spraying. Section 2 describes the physical/mathematical models used to describe the plasma flow, including turbulent and radiative transport models. Section 3 presents calculation procedures for the gas thermodynamic and transport properties that are fundamental for a realistic and accurate description of thermal plasmas. Boundary conditions, which seek to represent the physical phenomena dominating the interactions between the computational domain of the plasma flow and its surroundings, are described in Section 4. Section 5 describes models of the arc reattachment process, a complex phenomenon which severely alters the arc dynamics and is inherently present in the flow inside non-transferred arc torches. Finally, in Section 6, conclusions are drawn and some of the developments required for furthering the current understanding of plasma torch operation and achieving truly predictive DC arc plasma torch models, are presented.

## 2. Plasma Flow Models

### *2.1 Fluid Models*

The plasma formed in DC arc torches is an example of a thermal plasma. Among other things, thermal plasmas are characterized by high electron density (ranging between $10^{21}$ and $10^{24}$ $m^{-3}$) and high collision frequencies among its constituents (i.e. molecules, atoms, ions, and electrons) (Ref 1). High collision frequencies lead to a state close to local thermodynamic equilibrium (LTE) in which the kinetic energy of the constitutive species can be characterized by a single temperature. The LTE approximation is often violated near the plasma boundaries, either as the plasma interacts with solid walls, the working material, or with the cold working gas or atmosphere.

Thermal plasmas, due to their relatively high densities and pressures, and hence small mean free paths of the constitutive species, are most appropriately described by fluid models. Fluid models describe the evolution of the moments of the Boltzmann equation for each species in the plasma, and provide direct measures of macroscopic flow properties, such as temperature and pressure.

The fluid part of thermal plasma models can be expressed as a set of general transport equations expressed in conservative form as a balance among accumulation, net flux, and production; namely:

$$\underbrace{\frac{\partial \psi}{\partial t}}_{\text{accumulation}} + \underbrace{\nabla \cdot \mathbf{f}_{\psi}}_{\text{net flux}} - \underbrace{s_{\psi}}_{\text{production}} = 0 \,, \qquad \text{(Eq 1)}$$

where $\psi$ is a conserved property, $t$ represents time, $\mathbf{f}_{\psi}$ is the total (i.e. advective plus diffusive) flux of $\psi$, and $s_{\psi}$ is the net production/depletion rate of $\psi$.

**Table 1 Conservation equations of the thermodynamic and chemical equilibrium thermal plasma model**

| Conservation | Accumulation | Net flux | Net production |
|---|---|---|---|
| Total mass | $\frac{\partial \rho}{\partial t}$ | $\nabla \cdot (\mathbf{u}\rho)$ | 0 |
| Mass-ave. momentum | $\frac{\partial \rho \mathbf{u}}{\partial t}$ | $\nabla \cdot (\mathbf{u} \otimes \rho \mathbf{u} + p\boldsymbol{\delta} + \boldsymbol{\tau})$ | $\mathbf{J}_q \times \mathbf{B}$ |
| Internal energy | $\frac{\partial \rho h}{\partial t}$ | $\nabla \cdot (\mathbf{u}\rho h + \mathbf{q}')$ | $\frac{Dp}{Dt} - \boldsymbol{\tau} : \nabla \mathbf{u} + \mathbf{J}_q \cdot (\mathbf{E} + \mathbf{u} \times \mathbf{B}) - \dot{Q}_r$ |

**Table 2 Conservation equations of the thermodynamic and chemical non-equilibrium thermal plasma model**

| Conservation | Accumulation | Net flux | Net production |
|---|---|---|---|
| Total mass | $\frac{\partial \rho}{\partial t}$ | $\nabla \cdot (\mathbf{u}\rho)$ | 0 |
| Species mass | $\frac{\partial \rho_s}{\partial t}$ | $\nabla \cdot (\mathbf{u}\rho_s + \mathbf{J}_s)$ | $\dot{\rho}_s^c$ |
| Mass-ave. momentum | $\frac{\partial \rho \mathbf{u}}{\partial t}$ | $\nabla \cdot (\mathbf{u} \otimes \rho \mathbf{u} + p\boldsymbol{\delta} + \boldsymbol{\tau})$ | $\mathbf{J}_q \times \mathbf{B}$ |
| Int. energy heavy species | $\frac{\partial \rho h_h}{\partial t}$ | $\nabla \cdot (\mathbf{u}\rho h_h + \mathbf{q}'_h)$ | $\frac{Dp_h}{Dt} - \boldsymbol{\tau} : \nabla \mathbf{u} + \dot{Q}_{eh}$ |
| Int. energy electrons | $\frac{\partial \rho h_e}{\partial t}$ | $\nabla \cdot (\mathbf{u}\rho h_e + \mathbf{q}'_e)$ | $\frac{Dp_e}{Dt} + \mathbf{J}_q \cdot (\mathbf{E} + \mathbf{u} \times \mathbf{B}) - \dot{Q}_r - \dot{Q}_{eh}$ |

The most frequently used thermal plasma models rely on the LTE approximation, and model the plasma flow as the flow of a property-varying electromagnetic reactive fluid in chemical equilibrium in which the internal energy of the fluid is characterized by a single temperature $T$. The set of conservation equations describing such a flow is shown in Table 1. A more detailed description of a thermal plasma flow is given by allowing thermodynamic non-equilibrium (non-LTE or NLTE) between the electrons and the heavy species; that is, the internal energy of the fluid is now characterized by two temperatures: The electron temperature $T_e$ and the heavy species temperature $T_h$. Due to this fact, thermodynamic non-equilibrium thermal plasma models are also known as *two-temperature models*. The set of equations describing a NLTE thermal plasma in chemical non-equilibrium is listed in Table 2.

In Tables 1 and 2, $\rho$ represents the total mass density, $\rho_s$ the mass density of species $s$, $\mathbf{u}$ the mass-averaged velocity, $\mathbf{J}_s$ the mass diffusion flux and $\dot{\rho}_s^c$ the volumetric production rate of species $s$; $p$ represents the pressure, $\boldsymbol{\delta}$ the Kronecker delta, $\boldsymbol{\tau}$ the stress tensor, $\mathbf{J}_q$ the current density, $\mathbf{B}$ the magnetic field, $\mathbf{J}_q \times \mathbf{B}$ the Lorentz force; $h$, $h_h$, and $h_e$ the equilibrium, heavy-species and electron enthalpy, respectively (no subscript indicates an equilibrium or total property, while the subscripts $h$ and $e$ stand for heavy-particle and electron properties, respectively); $\mathbf{q}'$ the total heat flux; $Dp/Dt$ is the pressure work with $D/Dt$ as the substantive derivative; the term $\mathbf{J}_q \cdot (\mathbf{E} + \mathbf{u} \times \mathbf{B})$ represents the Joule heating, $\dot{Q}_r$ the volumetric net radiation losses, and $\dot{Q}_{eh}$ the electron – heavy-particle energy exchange term, which couples the two energy equations in the NLTE model. In Table 2, only $ns-1$, where $ns$ is the number of species, species mass conservation equations are required because the total mass conservation equation is included in the system.

Several assumptions and approximations are implied in the equations in Table 1 and Table 2. Particularly, closure of the moments of Boltzmann equation is taken into account in the specification of diffusive fluxes and/or transport coefficients. Furthermore, there are different forms to express the conservation equations in Tables 1 and 2, e.g. one could use conservation of total energy instead of internal energy. The most important requirement when formulating equilibrium or non-equilibrium plasma fluid models is self-consistency, which implies consistency with the moments of Boltzmann equation. In this regard, up to the specification of diffusive fluxes and source terms $\dot{\rho}_s^c$, $\dot{Q}_{eh}$, and $\dot{Q}_r$, the above models are self-consistent. Moreover, the LTE and NLTE models above are consistent with each other in the sense that the NLTE model gets reduced to the LTE model if thermal and chemical equilibrium are assumed (i.e. if one enforces $T_h = T_e$ in the equations in Table 2 and if the plasma composition is determined only as function of the thermodynamic state of the fluid). Furthermore, the addition of the electron and heavy-species energy equations in Table 2 produces the total internal energy conservation equation in Table 1.

### *2.2 Diffusion Fluxes and Source Terms*

The systems of equations in Tables 1 and 2 are closed with the specification of diffusive fluxes $\mathbf{J}_s$, $\boldsymbol{\tau}$ and $\mathbf{q}'$, and the source terms $\dot{\rho}_s^c$, $\dot{Q}_{eh}$, and $\dot{Q}_r$.

The mass diffusion flux of species $s$ is given by:

$$\mathbf{J}_s = \rho_s(\mathbf{u}_s - \mathbf{u}), \tag{Eq 2}$$

where $\mathbf{u}_s$ is the species $s$ velocity. The evolution of $\mathbf{u}_s$ is described by a momentum conservation equation derived from the Boltzmann equation for species $s$. For a more rigorous treatment of chemical non-equilibrium than the one presented in Table 2, one momentum conservation equation should be solved for each species (Ref 2). This procedure is exceedingly expensive, especially for the modeling of industrial thermal plasma flows, as it would add ($3ns - 1$) equations. Thus, alternative approaches are sought. These approaches seek to define mass diffusion fluxes as function of the other macroscopic characteristics of the flow, such as temperature, pressure and concentration gradients.

The derivation of consistent mass diffusion fluxes for thermal plasmas is quite involved, especially for two-temperature plasmas, due to the transport of charged species coupled to the electromagnetic driving forces (Ref 3). One well-known approach for mass diffusion modeling in thermal plasmas is the Self-Consistent Effective Binary Diffusion (SCEBD) approximation of Ramshaw and Chang (Ref 4, 5). The SCEBD approximation models the mass diffusion fluxes according to:

$$\mathbf{J}_s = -\frac{D_s}{R_s T_s}\mathbf{G}'_s + \frac{\rho_s}{\rho}\sum_{j \neq s}\frac{D_{sj}}{R_j T_j}\mathbf{G}'_j, \tag{Eq 3}$$

where $D_s$ is the effective diffusivity of species $s$, $D_{sj}$ the binary diffusion coefficient between species $s$ and $j$, $R_s$ and $T_s$ are the gas constant and temperature of species $s$, respectively; and $\mathbf{G}'_s$ is the total driving force acting over species $s$, which is a function of the gradients of temperature, pressure, concentrations, and of external forces (electromagnetic and gravitational). Ramshaw and Chang's model is still relatively expensive to apply in simulations of industrial thermal plasma processes. A more practical approach is the Combined Diffusion method developed by Murphy (Ref 6) and extended by Rat *et al* (Ref 7). This approach is based on the definition of combined diffusion coefficients together with the grouping of species according to their parent gases. This model allows the description of a thermal plasma in chemical and thermodynamic non-equilibrium by the transport of groups of species related by their parent gases. As an example, an Ar-He plasma in chemical non-equilibrium can be modeled by conservation equations of the group of species related to Ar (that is, Ar, $Ar^+$, $Ar^{++}$) and the group related to He (He, $He^+$). This approach is valid if the parent gases (e.g. Ar and He above) do not react with each other.

The diffusive transport of momentum is modeled by the stress tensor $\boldsymbol{\tau}$, which is defined as for a Newtonian fluid and is given by:

$$\boldsymbol{\tau} = -\mu(\nabla\mathbf{u} + \nabla\mathbf{u}^t - \tfrac{2}{3}(\nabla\cdot\mathbf{u})\boldsymbol{\delta}), \qquad \text{(Eq 4)}$$

where $\mu$ is the dynamic viscosity, the superscript $^t$ indicates the transpose of matrix $\mathbf{u}$, and the 2/3 factor in the fluid dilatation $\nabla\cdot\mathbf{u}$ comes from the Stoke's hypothesis for the dilatational viscosity.

The total heat fluxes in the LTE and NLTE models describe the heat transported by conduction and the enthalpy transport by mass diffusion. They are defined by:

$$\mathbf{q}' = -\kappa\nabla T + \sum_s h_s\mathbf{J}_s, \qquad \text{(Eq 5)}$$

$$\mathbf{q}'_h = -\kappa_h\nabla T_h + \sum_{s\neq e} h_s\mathbf{J}_s, \qquad \text{(Eq 6)}$$

$$\mathbf{q}'_e = -\kappa_e\nabla T_e + h_e\mathbf{J}_e, \qquad \text{(Eq 7)}$$

where the summation in Eq 5 runs over all the species in the plasma; and $\kappa$, $\kappa_h$ and $\kappa_e$ are the total, heavy species, and electron translational thermal conductivities, respectively.

A very useful and widely spread practice used in chemical-equilibrium thermal plasma models is the use of a reactive thermal conductivity. Given that chemical equilibrium is assumed, the plasma composition is only a function of the thermodynamic state of the plasma (e.g. the plasma composition can be expressed as function of the equilibrium temperature $T$ and total pressure $p$). Using this approximation, the mass diffusion fluxes of heavy species ($\mathbf{J}_s$) can be expressed as functions of $\nabla T$ and $\nabla p$. Neglecting the contribution due to the pressure gradient, the total heat transferred by conduction plus the heat transported by heavy species can be expressed as a reactive thermal conductivity $\kappa_r$ times $\nabla T$; namely:

$$-\kappa\nabla T + \sum_{s\neq e} h_s \mathbf{J}_s = -\kappa_r \nabla T\,. \qquad \text{(Eq 8)}$$

The reactive thermal conductivity can be treated as any other transport property. It is primary a strong function of the temperature $T$ and only weakly a function of pressure. The energy transported by electron mass diffusion in Eq 7 can be approximated by:

$$\mathbf{J}_e \approx -\frac{m_e}{e}\mathbf{J}_q\,, \qquad \text{(Eq 9)}$$

where $e$ is the elementary electric charge and $m_e$ is the electron mass. Equation 9 neglects the charge transported by the heavy species, which is a valid approximation for most thermal plasmas. The mass diffusion flux of electrons is not included in the definition of the reactive thermal conductivity because, as clearly implied in Eq 9, this flux is mostly driven by the electrical characteristics of the system. By introducing Eq 8 and Eq 9 in Eq 5, we obtain the final expression for the total heat flux which is often found in the literature, namely:

$$\mathbf{q}' = -\kappa_r \nabla T - \frac{m_e}{e}\mathbf{J}_q\,. \qquad \text{(Eq 10)}$$

Equation 10 is particularly useful because it has a simple form and explicitly expresses the main factors driving the transport of heat through the plasma.

The species production term $\dot{\rho}_s^c$ is similar to that found in standard reactive fluid dynamics literature (e.g. Ref 8) and is given by:

$$\dot{\rho}_s^c = M_s \sum_{r=1}^{nr} (b_{s,r} - a_{s,r})\dot{\varpi}_r\,, \qquad \text{(Eq 11)}$$

$$\dot{\varpi}_r = -k_{f,r}\prod_i^{ns}\left(\frac{\rho_i}{M_i}\right)^{a_{i,r}} + k_{b,r}\prod_i^{ns}\left(\frac{\rho_i}{M_i}\right)^{b_{i,r}}\,, \qquad \text{(Eq 12)}$$

where $M_s$ is the molecular weight of species $s$, $nr$ is the number of reactions, $\dot{\varpi}_r$ is the progress rate of reaction $r$, $a_{s,r}$ and $b_{s,r}$ are the reaction (e.g. stoichiometric) coefficients in the forward and backward directions for the reaction $r$, and $k_{f,r}$ and $k_{b,r}$ are the forward and backward reaction rates for reaction $r$.

The description of source term due to radiation transport is rather complex and will be explained in Section 2.4. The final term to close the NLTE model is the electron - heavy-species energy exchange term $\dot{Q}_{eh}$ which is often modeled as (e.g. Ref 9):

$$\dot{Q}_{eh} = \sum_{s \neq e} \frac{3}{2} k_B \frac{2 m_s m_e}{(m_s + m_e)^2} \nu_{es} \delta_{es} (T_e - T_h), \qquad \text{(Eq 13)}$$

where $k_B$ is the Boltzmann constant, $\nu_{es}$ is the collision frequency between electrons and species $s$, and $\delta_{es}$ is the inelastic collision factor, which is equal to 1 for atomic species. The term $\dot{Q}_{eh}$ models the average exchange of kinetic energy per unit volume between electrons and heavy species.

## *2.3 Electromagnetic Equations*

The fluid models described in Section 2.1 are complemented with the equations describing the evolution of the electromagnetic fields. These equations are the Maxwell's equations, which, for typical thermal plasmas, are simplified by neglecting relativistic effects, magnetization, as well as charge accumulation. They are listed in Table 3.

**Table 3 Maxwell's equations for thermal plasmas**

| Name | Equation |
|---|---|
| Ampere's law: | $\nabla \times \mathbf{B} = \mu_0 \mathbf{J}_q$ |
| Faraday's law: | $\nabla \times \mathbf{E}_p = -\frac{\partial \mathbf{B}}{\partial t}$ |
| (*Generalized*) Ohm's law: | $\mathbf{J}_q = \sigma(\mathbf{E}_p + \mathbf{u} \times \mathbf{B})$ |
| Gauss' law (charge conservation): | $\nabla \cdot \mathbf{J}_q = 0$ |
| Solenoidal Constraint: | $\nabla \cdot \mathbf{B} = 0$ |

In Table 3, $\mu_0$ represents the permeability of free space, $\sigma$ electrical conductivity, and $\mathbf{E}_p$ the effective electric field. The latter is used in Table 3, instead of the *real* electric field **E**, in order to account for the so-called *generalized* Ohm laws. These laws take into account the dynamic modification of the electromagnetic fields due to charge transport (i.e. charge transport is implied by the mass diffusion fluxes $\mathbf{J}_s$ of charged species) and, therefore, need to be consistent with the mass diffusion model used in the fluid formulation (e.g. Eq 3). For LTE chemical-equilibrium models, it is often assumed that $\mathbf{E} = \mathbf{E}_p$. For NLTE models, the main modification of the electric field is due to the electron pressure gradient as shown in the following expression:

$$\mathbf{E}_p \approx \mathbf{E} + \frac{\nabla p_e}{e n_e}, \qquad \text{(Eq 14)}$$

where $n_e$ is the electron number density. In more complete generalized Ohm laws, the effective electric field is given by nonlinear expressions (e.g. $\mathbf{E}_p$ is a function of $\mathbf{J}_q$x$\mathbf{B}$). The reader may realize that the Joule heating term in Table 2 involves the real electric field **E** and not the effective one $\mathbf{E}_p$; an interesting discussion of the derivation of this term is found in (Ref 10).

The Maxwell's equations listed in Table 3 can be expressed in different forms. Particularly useful for thermal plasma flow solvers are the expressions based on electromagnetic potentials

$$\mathbf{E}_p = -\nabla\phi_p - \frac{\partial \mathbf{A}}{\partial t} \text{ and} \qquad \text{(Eq 15)}$$

$$\nabla \times \mathbf{A} = \mathbf{B}, \qquad \text{(Eq 16)}$$

where $\phi_p$ is the *effective* electric potential (usually assumed equal to the electric potential $\phi$ in LTE models) and **A** the magnetic vector potential. The use of the magnetic potential has the added advantage that the solenoidal constraint is satisfied *a priori*. Using these potentials, Maxwell's equations can be expressed by:

$$\frac{\partial \mathbf{A}}{\partial t} + \nabla\phi_p - \mathbf{u} \times (\nabla \times \mathbf{A}) - \frac{1}{\mu_0 \sigma}\nabla^2 \mathbf{A} = \mathbf{0} \text{ and} \qquad \text{(Eq 17)}$$

$$\nabla \cdot \sigma(\nabla\phi_p + \frac{\partial \mathbf{A}}{\partial t} - \mathbf{u} \times \nabla \times \mathbf{A}) = 0. \qquad \text{(Eq 18)}$$

Equation 17 is commonly known as a form of magnetic induction equation, whereas Eq 18 is just an expression of charge conservation. Equation 17 can alternatively be expressed as:

$$\nabla^2 \mathbf{A} = -\mu_0 \mathbf{J}_q \,. \qquad \text{(Eq 19)}$$

Equation 19 is the most often used form of induction equation used in the thermal plasma modeling literature, particularly as it is expressed as a relatively simple diffusion equation and therefore amenable for its solution in fluid flow solvers.

## *2.4 Turbulence Models*

In DC arc plasma torches the working gas is typically at ambient temperature when it enters the torch. The temperature of the gas, as it interacts with the arc, increases by a rate in the order of $10^4$ K/mm. This rapid heating causes the sudden expansion of the gas and consequently its rapid acceleration. The velocity of the gas across the torch often varies by 2 orders of magnitude (e.g. from $O(10)$ to $O(1000)$ m/s). The large gas acceleration and shear velocity and temperature gradients inside the torch, together with the electromagnetic forcing (Ref 11, 12), cause the flow to become unstable and turbulent. Turbulence is further enhanced when the plasma flow leaves the torch and interacts with the cold and, thus, denser environment.

The accurate modeling of turbulent flows, due to their large range of length and time scales, represents a great challenge. The most faithful numerical description of turbulent flows is given by the approach known as Direct Numerical Simulation (DNS), which seeks to resolve all the scales of the flow without any approximation (i.e. by definition, no physical, e.g. eddy viscosity, see below, or numerical, e.g. upwinding, dissipation mechanisms are employed). DNS of large Reynolds number (*Re*) flows and for industrial geometries are almost unfeasible with current computing power, as the range of length scales to be solved (i.e. the size of the grid needed) scales as $Re^3$. DNS of non-equilibrium plasmas can be found in the literature (e.g. Ref 13) but, to the best knowledge of the authors, no DNS of a thermal plasma has been performed to date.

The large cost of DNS has motivated a variety of alternative approaches to simulate turbulent flows. The main approaches are grouped in what are known as Large Eddy Simulations (LES), which seek to model only the small scales of the flow, and Reynolds-Averaged Navier-Stokes (RANS) models, which seek the solution of approximations of the time-averaged Navier-Stokes Equations. LES is typically far more accurate than RANS, but often an order of magnitude or more more expensive. Recently, the technique known as Detached Eddy Simulation (DES) has been gaining more acceptance, especially in the modeling of external flows, as it mixes the LES and RANS approaches: LES is used where it is most needed (e.g. in highly unsteady zones) and RANS in the rest of the domain or where the use of LES would be prohibitive (e.g. near walls, where vorticity is mostly created).

Most LES and RANS models rely on the Boussinesq hypothesis, which consists of modeling the turbulent stresses in a similar manner as the viscous stress and, hence, reduces the formulation of the turbulence model to the specification of an appropriate *turbulent viscosity* $\mu_t$ (the total stress $\boldsymbol{\tau}$ is still given by Eq 4 but $\mu$ is replaced with $\mu + \mu_t$). More sophisticated models exist (Ref 14), which seek to model the whole turbulent stress with very few or none empirical approximations, such as the residual-driven model of Bazilevs *et al* (Ref 15). But these models, although they are potentially the best approaches for the modeling of complex turbulent flows, are not widely used yet.

Diverse RANS and few LES models are often offered in commercial Computational Fluid Dynamics (CFD) software, which has driven the increasing use of these models. Turbulent viscosity models for LES are usually far simpler than models for RANS. But, sound LES simulations require highly accurate spatial and temporal discretizations. These requirements are usually hard to satisfy using commercial software because robustness, probably the most important feature in commercial software, is very often in opposition to accuracy (e.g. it is harder to obtain convergence using more accurate, high order, discretizations).

The use of turbulence models in thermal plasma flows is significantly more involved than for most other industrial applications due to their inherent characteristics (i.e. reactivity, large property variations and electromagnetic effects). The use of standard turbulent models for thermal plasma simulation often implicitly neglects several of these characteristics.

Although some LES of DC arc plasma torches have been performed (Ref 16, 17), by far RANS simulations dominate the thermal plasma literature. RANS models range from zero, one, and two equation models to Reynolds-stress models, which model each component of the turbulent stress tensor. By far the most widely used turbulence model in thermal plasma modeling is the $k$-$\varepsilon$ model developed by Launder and Spalding (Ref 18), where $k$ stands for the turbulent kinetic energy and $\varepsilon$ its rate of dissipation (Table 4). In the derivation of the standard $k$-$\varepsilon$ model, the flow is assumed to be fully turbulent, and the effects of molecular viscosity are negligible. Furthermore, the standard $k$-$\varepsilon$ is a semi-empirical model, and the derivation of the model equations relies on phenomenological considerations and empiricism.

**Table 4 Equations of the standard k-ε turbulent model**

| Cons. | Accumulation | Net flux | Net production |
|---|---|---|---|
| Turbulent kinetic energy | $\frac{\partial \rho k}{\partial t}$ | $\nabla \cdot (\mathbf{u}\rho k - (\mu + \frac{\mu_t}{\sigma_k})\nabla k)$ | $G_k - \rho\varepsilon$ |
| Rate of dissipation | $\frac{\partial \rho\varepsilon}{\partial t}$ | $\nabla \cdot (\mathbf{u}\rho\varepsilon - (\mu + \frac{\mu_t}{\sigma_\varepsilon})\nabla \varepsilon)$ | $c_{\varepsilon 1} G_k \frac{\varepsilon}{k} - c_{\varepsilon 2}\rho \frac{\varepsilon^2}{k}$ |

In Table 4, $\sigma_k$ and $\sigma_\varepsilon$ are the Prandtl numbers for $k$ and $\varepsilon$ respectively, $G_k$ represents the generation of turbulent kinetic energy due to the mean velocity gradients, and $c_{\varepsilon 1}$ and $c_{\varepsilon 2}$ are model constants. The turbulent viscosity for the $k$-$\varepsilon$ model is defined by:

$$\mu_t = c_\mu \rho \frac{k^2}{\varepsilon} \quad \text{(Eq 20)}$$

More advanced $k$-$\varepsilon$ models have been developed, which are more rigorously derived, more accurate. They are valid for a wider variety of flows, and are also often available in several commercial CFD software, like the RNG and the *realizable* $k$-$\varepsilon$ models available in Fluent (Ref 19). For example, the standard $k$-$\varepsilon$ is only valid for fully turbulent flows, while the RNG $k$-$\varepsilon$ is valid for fully turbulent as well as for low- Reynolds number and near-wall flows.

To summarize, the use of turbulence models for the modeling of the flow in DC arc torches has to be approached with care and weighting the assumptions and approximations involved. Particularly, for the flow in non-transferred torches, which is highly unsteady, a LES approach is more appropriate; whereas for the modeling of transferred torches, especially when the flow is steady, the use of RNG $k$-$\varepsilon$ or similar models may provide an adequate description of the flow. Nevertheless, validation with experimental measurements is required, and when possible should be pursued.

## *2.5 Radiative Transport*

Radiative transfer in thermal plasmas involves line and continuum radiation, including bremsstrahlung and recombination radiation (Ref 1).

The total radiative flux source term of the energy equations shown in Tables 1 and 2 is given by:

$$\dot{Q}_r = \nabla \cdot \mathbf{q}_r , \quad \text{(Eq 21)}$$

where $\mathbf{q}_r$ represents the radiative heat flux. The net radiation flux is a function of the spectral intensity $I_\lambda(\mathbf{x}, \mathbf{s})$, which measures the radiation intensity in the point $\mathbf{x}$ along the direction $\mathbf{s}$ for a given wavelength $\lambda$, according to:

$$\nabla \cdot \mathbf{q}_r = 4\pi \int_0^\infty \kappa_\lambda I_{b\lambda} d\lambda - \int_0^\infty \int_0^{4\pi} \kappa_\lambda I_\lambda d\Omega d\lambda , \quad \text{(Eq 22)}$$

with $I_{b\lambda}$ as the spectral black body intensity, $\kappa_\lambda$ the spectral absorption coefficient (which is a function of the gas composition, pressure, temperature(s) and wavelength), and $\Omega$ the solid angle. The second term on the right hand of Eq 22 implies integration over all directions and wavelengths. From Eq 22, the radiative source term represents the difference between the emission and absorption occurring at a given location $\mathbf{x}$. The spectral intensity $I_\lambda$ is found by solving the Radiative Transfer Equation (RTE), which after neglecting time-dependency, scattering, and refraction effects, can be expressed as:

$$\mathbf{s} \cdot \nabla I_\lambda(\mathbf{x}, \mathbf{s}) = \kappa_\lambda (I_{b\lambda} - I_\lambda) . \quad \text{(Eq 23)}$$

Equation 23 models the energy transported by photons through the flow. Radiative media can often be characterized by their *optical thickness*, which is a measure of the interaction that the photons experience as they travel through a domain. It can be estimated by $\kappa_\lambda L$, where $L$ is a characteristic length of the domain (e.g. torch diameter). In this regard, the plasma flow in DC arc torches is usually considered as optically thin ($\kappa_\lambda L << 1$) because it is often assumed that the photons leave the plasma with very little interaction with the flow.

The direct solution of the RTE is exceeding expensive due to the dual **s** to **x** dependence, and consequently diverse types of approximations are often employed. The detailed description of the radiative transport in thermal plasmas represents an enormous challenge not only because of the complex absorption spectra of the species present, but, also due to the weak interaction of the photons with the surrounding media. This last characteristic jeopardizes the use of models that rely on strong coupling (optically thick media), like diffusion-like models such as the P1 approximation, and makes mandatory the use of more computationally expensive techniques like Direct Simulation Monte Carlo or directional transport methods, like ray-tracing techniques and Discrete Ordinates Methods (DOMs). The DOM consists of solving the RTE along few ordinate directions transforming the RTE in a (typically large) set of transport equations. The P1 approximation consists of the formulation of a transport equation (of reaction-diffusion form) for the first order term of the expansion of the radiative intensity in spherical harmonics. The P1 method is a good approximation of the radiation transport in optically thick media, and hence is not suitable for most DC arc plasma torch modeling.

Probably one of the best radiation transfer simulations applied to a thermal plasma flow is the work of Menart *et al* (Ref 20) who used a DOM for a large set of wavelengths. Because their work was focused on analyzing the radiative transfer, Menart *et al* did not solve the radiative transport coupled to a plasma flow model. Instead, they used a pre-calculated temperature field in order to determine $\kappa_\lambda(T, \lambda)$ to solve the RTE. Their approach is justified by the enormous computational cost required to solve the plasma flow together with radiative transport. More recently, Iordanidis and co-workers (Ref 21, 22) compared the DOM and P1 methods and performed simulations of the plasma flow in circuit breakers using the DOM due to its greater accuracy. An alternative approach is the use of view factors to determine the exchange of radiative energy among the domain boundaries. Such approach has been successfully used by Lago *et al* (Ref 23) for the simulation of a free-burning arc.

The form in which radiation directly interacts with the plasma flow (namely Eq 21) suggests that detailed description of radiation transfer may not be needed and that direct approximations of $\dot{Q}_r$ could be used. This term can be approximated following a classical approach as done by Tanaka in (Ref 24) for the modeling of an inductively coupled torch. But, by far, the most common approximation used in thermal plasma modeling is the use of the effective net emission approximation (Ref 25, 26, 27). Under this approximation, the net radiative flux is approximated according to:

$$\dot{Q}_r = 4\pi\varepsilon_r \qquad \text{(Eq 24)}$$

where $\varepsilon_r$ is the effective net emission coefficient, which for a given plasma forming gas can be expressed as function of temperature(s) and an effective absorption radius $R_r$. The latter represents the radius of a sphere in which the emitted radiation can be re-absorbed; outside of this sphere, the emitted radiation leaves without further interaction with the plasma (hence, the optically-thin approximation implies $R_r$ = 0). The net emission approach is particularly appealing for thermal plasma flow simulation because the effective emission coefficient can be treated as any other thermodynamic or transport property of the plasma.

At present, net emission coefficient (NEC) are available for the following pure gases (Ar, $O_2$, $H_2O$, Air) and mixtures (Ar–$H_2$, Ar-Fe, Ar-Fe-$H_2$, Air-metallic vapors) (e.g. see Ref 28, 29, 30, 31, 32).

# 3 Gas Properties

In plasma simulations, mass, momentum and energy equations, together with electromagnetic field equations (see Section 2.1), have to be solved with a coupled approach and the accuracy of results depends strongly on the use of suitable thermodynamic and transport properties (see Section 2.2).

## *3.1 LTE thermodynamic and transport properties*

The determination of thermodynamic and transport properties requires first the calculation of plasma composition that can be obtained either from a chemical non-equilibrium model (e.g. see Table 2) or chemical equilibrium models based on mass action laws or minimization of Gibbs free energies. Thermodynamic properties are directly calculated from the particle number densities of the various species forming the plasma and previous knowledge of the internal partition functions.

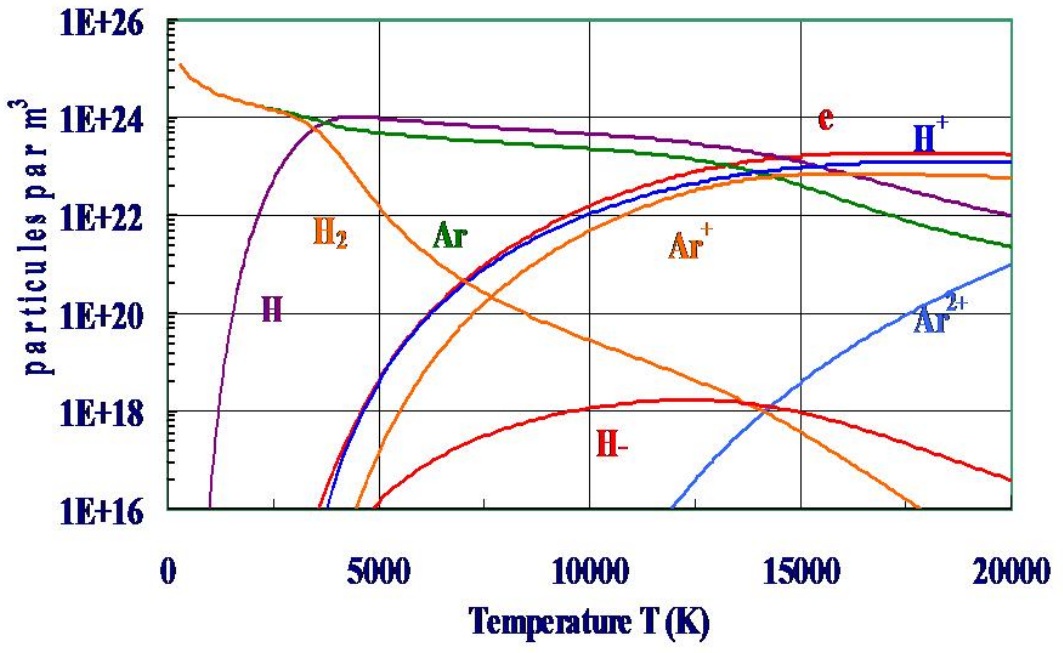

(a)

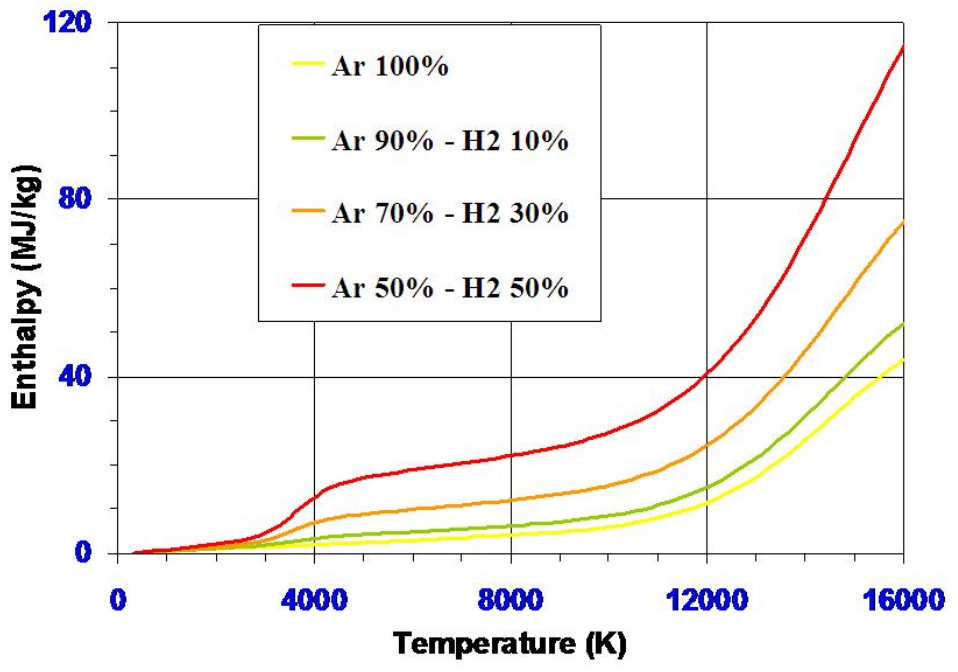

(b)

**Fig. 3** (a) Temperature dependence of the composition (species number densities) in an argon-hydrogen (75/25% vol) at atmospheric pressure and (b) specific enthalpy (MJ/kg) of various Ar-$H_2$ gas mixture at atmospheric pressure versus temperature (Ref 33)

Figure 3 shows the composition of an Ar-$H_2$ mixture; a common gas mixture used in plasma spraying because the addition of hydrogen to the plasma gas mixture increases both the specific enthalpy and thermal conductivity of the plasma flow, especially at temperatures where dissociation and ionisation occur. The tendencies of the curves are explained by the lack of reactions between Ar and $H_2$. The evolution of specific enthalpy is also shown in Fig. 3 (b). It can be seen that the addition of hydrogen to argon increases the specific enthalpy of the mixture. Nevertheless, due to the high molar mass of argon compared to that of hydrogen, the increase in specific enthalpy becomes significant when the hydrogen content is higher than 30%. For more complex mixtures (e.g. Ar-$H_2$ and air from the surrounding atmosphere, in plasma spraying) mixing rules are often used (Ref 34).

Once the composition is known, the computation of heat, mass and momentum fluxes (see Section 2.2) requires the knowledge of transport properties (see Tables 1 and 2). The calculation is based on solving the Boltzmann integro-differential equation describing the evolution of the electron energy distribution function (EEDF), by the Chapman–Enskog (CE) method (Ref 35) applied to complex mixtures. This methodology has been analysed exhaustively by Hirschfelder *et al* (Ref 36). Although established for weakly ionized gases, this method has been demonstrated to be valid for thermal plasmas (Ref 37). The distribution function for different species is assumed to be Maxwellian with a first-order perturbation function which is developed in the form of a series of Sonine polynomials: This reduces the Boltzmann equation to a set of linear equations whose solution makes it possible to obtain the gas transport properties. The coefficients of the set of linear equations depend on collision integrals which take into account the binary interaction between colliding species. The computation of these data requires the knowledge of either the interaction potential, which describes the collision dynamics, or the transport cross-sections, which can be derived from differential cross-sections, quantum phase shifts or experimental data.

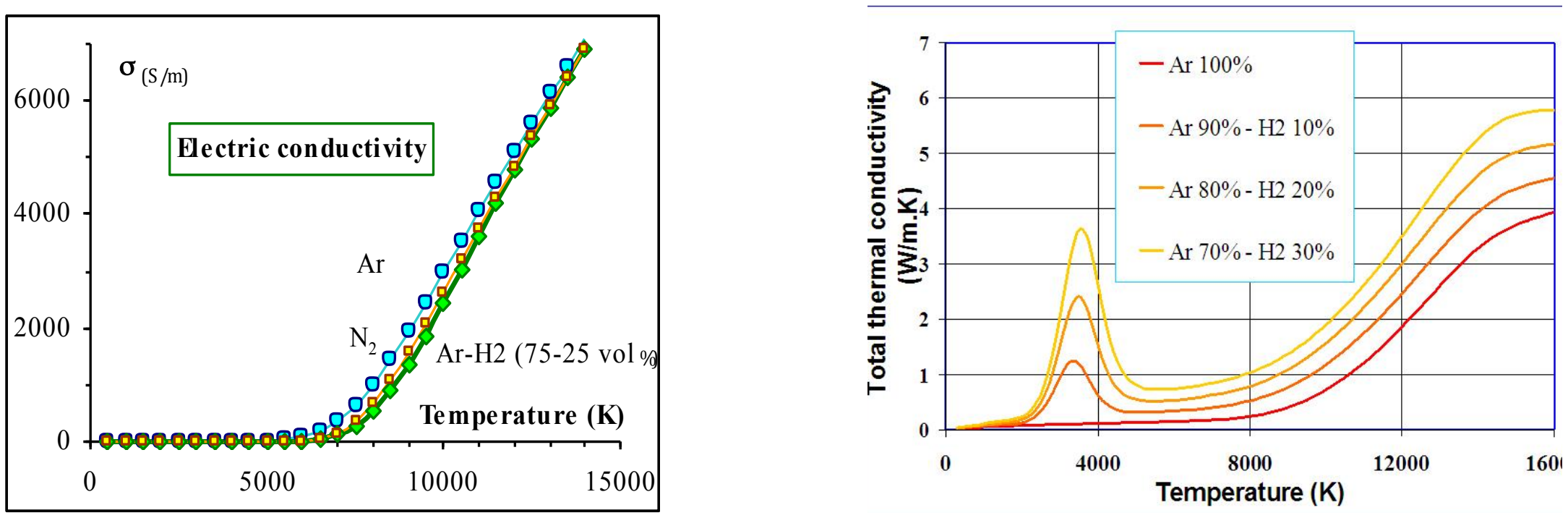

(a) (b)

**Fig. 4** Temperature dependence of the electrical conductivity of Ar, Ar-$H_2$ and $N_2$ gas (a) and thermal conductivity of an Ar-$H_2$ gas mixtures at atmospheric pressure (b)

Viscosity and electrical conductivity are obtained by the direct method (Ref 35) using different orders of approximation (Ref 1). The evolution of the electrical conductivity $\sigma$ versus temperature $T$ (Fig. 4 (a)) for various gases, shows very similar temperature dependencies, with a critical temperature, $T_c$, about 7000 K, under which $\sigma$ is vanishing to zero and above which $\sigma$ increases linearly up to 14000 K.

Thermal conductivity (see Fig. 4 (b)) is written as the sum of four components (see Eq. 5 in Section 2.2): one term due to the translation of heavy particles, a second due to the translation of the electrons, a third representing the internal thermal conductivity and the last term corresponding to the reaction thermal conductivity (Ref 38). When $H_2$ is added to Ar, the thermal conductivity of the mixture increases with the percentage of $H_2$, especially near the dissociation and ionisation temperatures due to the reactive contribution as illustrated in Fig. 4 (b).

The numerical treatment of mass diffusion is particularly complicated, since a large number of diffusion coefficients ($ns(ns-1)/2$ ordinary diffusion coefficients and $ns-1$ thermal diffusion coefficients for $ns$ species) has to be considered. To simplify this calculation, Murphy (Ref 6) introduced the treatment of diffusion in terms of gases instead of species (for example, argon and hydrogen gases, instead of considering Ar, $Ar^+$, $Ar^{2+}$, $H_2$, H, $H^+$ and $e^-$). Total diffusion coefficients taking into account ambipolar diffusion and temperature or pressure gradients have been proposed by Devoto (Ref 39). The combined diffusion coefficients, very useful for gas mixture modelling, have been computed by Murphy (Ref 40).

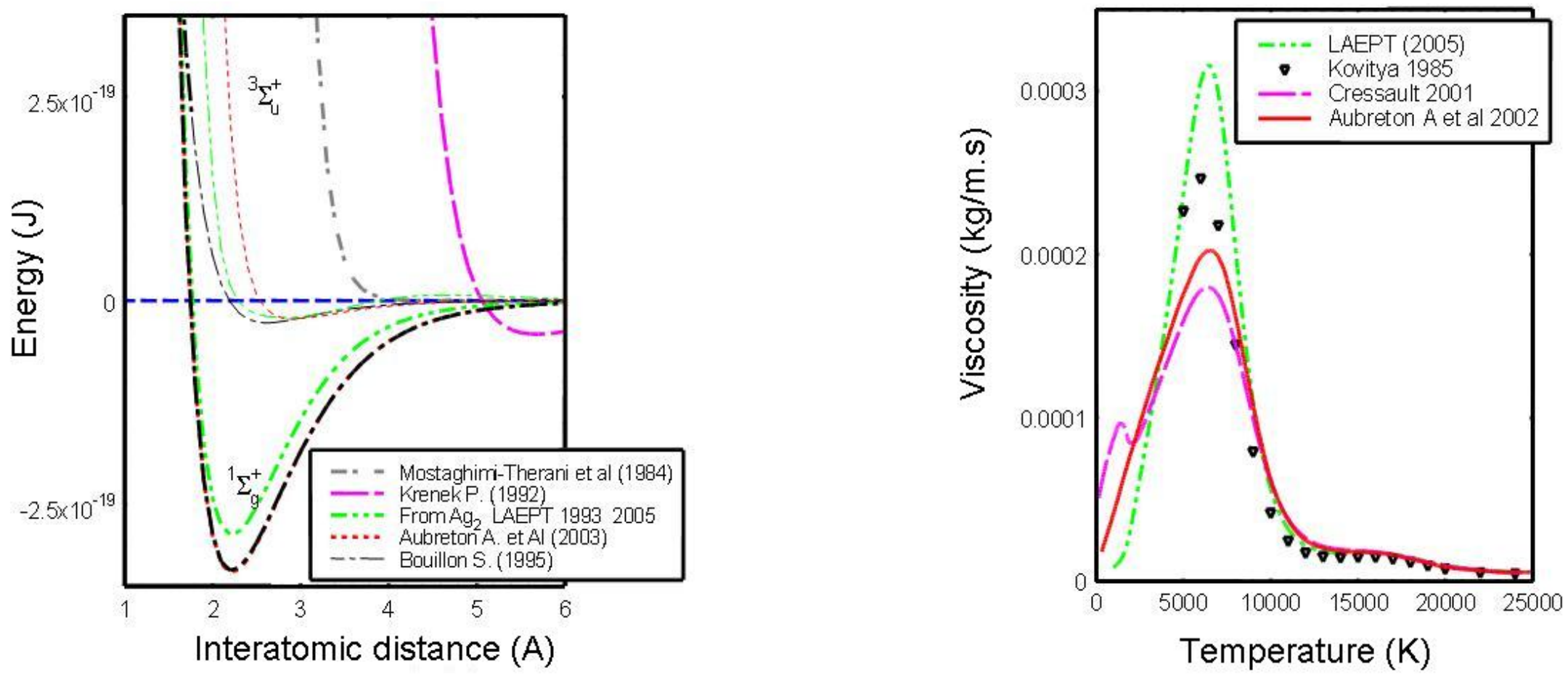

**Fig. 5** Influence of the choice of the Cu-Cu potential interaction on copper vapor viscosity

A large volume of data has been published for the thermodynamic and transport properties of gases. For thermal plasma-based processes, transport properties under LTE assumptions are available for the most used plasma gases (Ar, $H_2$, $N_2$, $O_2$, $H_2O$, He, $SF_6$, $CH_4$, air and their binary and ternary mixtures: Ar–$H_2$, Ar–He, Ar-Fe, Ar-$O_2$, Ar-$N_2$, air-$N_2$, air-Ar, air-$O_2$, air-$CH_4$, Ar–$H_2$-He, Ar-$H_2$-Cu etc. (Ref 41, 42, 43, 44, 45, 46, 47, 48, 49)). However, the use of these data, especially for transport properties, requires caution because the collision integrals that are the basis of calculations are not always well known, and this may lead to large uncertainties. Figure 5 shows, as an example, the effect of such uncertainties on the Cu-Cu interactions potential for the viscosity of copper vapor. Errors of a factor of 2 can occur in the viscosity due to the uncertainties in interaction potentials. The same uncertainties may occur in the calculation of other transport properties (Ref 1).

### *3.2 NLTE thermodynamic and transport properties*

Despite the usefulness of the LTE assumption, one must realize that deviation from LTE is much more the rule than the exception in plasma-based processes. For example, deviations from LTE occur close to the electrodes of the electric arc or in the boundary layer insulating electrically the arc column from the anode wall of a plasma spraying torch. Deviation from LTE also occurs when liquid or solid precursors are injected inside the plasma jet to treat powders or coatings. In that case, the energy distribution function (EEDF) of each kind of particles remains Maxwellian but the mean kinetic energy may be different for the electrons and heavy particles. The electron temperature $T_e$ is, then, higher than the heavy particle temperature $T_h$, and departure from thermal equilibrium may be characterized by the parameter $\theta$ defined as $\theta = T_e/T_h$.

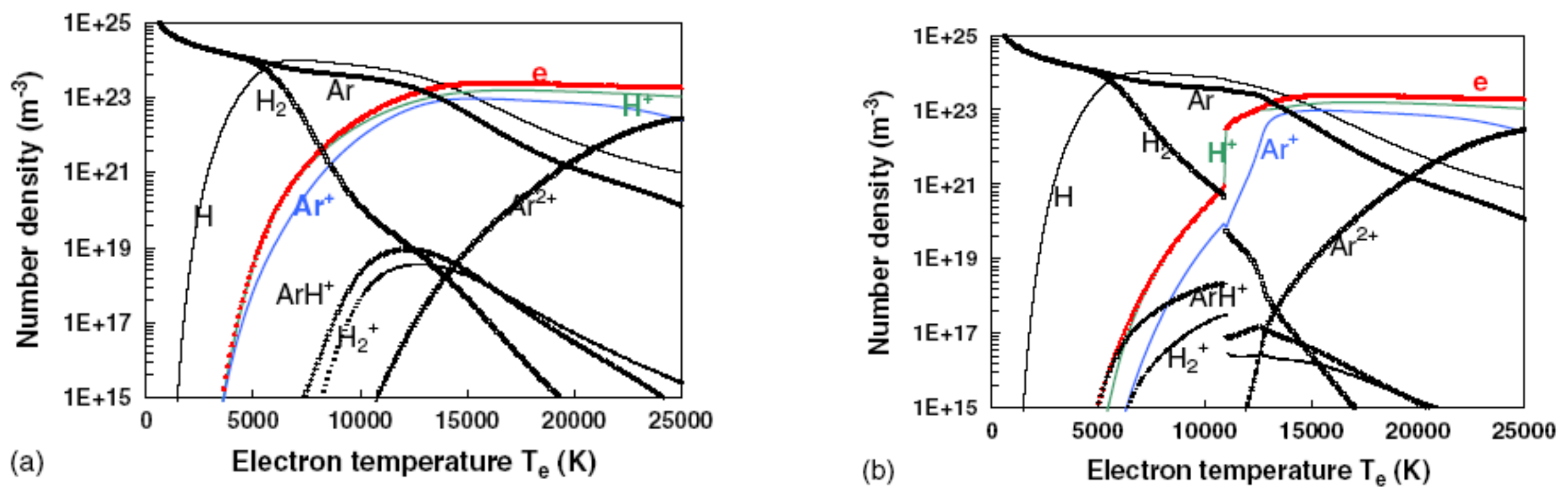

**Fig. 6** Dependence of the non equilibrium composition of an Ar-$H_2$ (50% vol) mixture on the electron temperature at atmospheric pressure for $\theta$ = 1.6 obtained using (a) Van der Sanden *et al*'s method and (b) kinetic calculation (Ref 51)

Similarly to LTE assumptions, calculation of the thermodynamic and transport properties requires prior calculation of the two-temperature plasma composition. Nevertheless, generalization of the mass action law and/or Gibbs free energy minimization to NLTE plasma has a long and tumultuous history as evidenced by the different approaches found in the literature (Ref 50).

Although all the methods developed converge at thermal equilibrium, they give different results when non-LTE is assumed. These discrepancies are highlighted in Fig. 6 (Ref 52) that presents a comparison between the results obtained from van de Sanden *et al*'s method and a steady-state kinetic calculation at chemical equilibrium for a non-equilibrium Ar-$H_2$ (50 mol%) plasma. While both methods (Fig. 6 (a) and (b)) give approximately the same results below 5000 K and above 15000 K, they strongly differ at intermediate temperatures. The main difference between the results is a discontinuity appearing at around $T_e$ = 11 000 K, which was also observed by Cliteur *et al* (Ref 53). Since no agreement can be found between the different theoretical approaches, it is apparent that only experimental measurements will be helpful to validate the above results. Unfortunately, the calculations are in general not backed by experimental validations, and are generally very difficult to compare with measurements reported in literature since the conditions of calculation do not often match with experimental conditions.

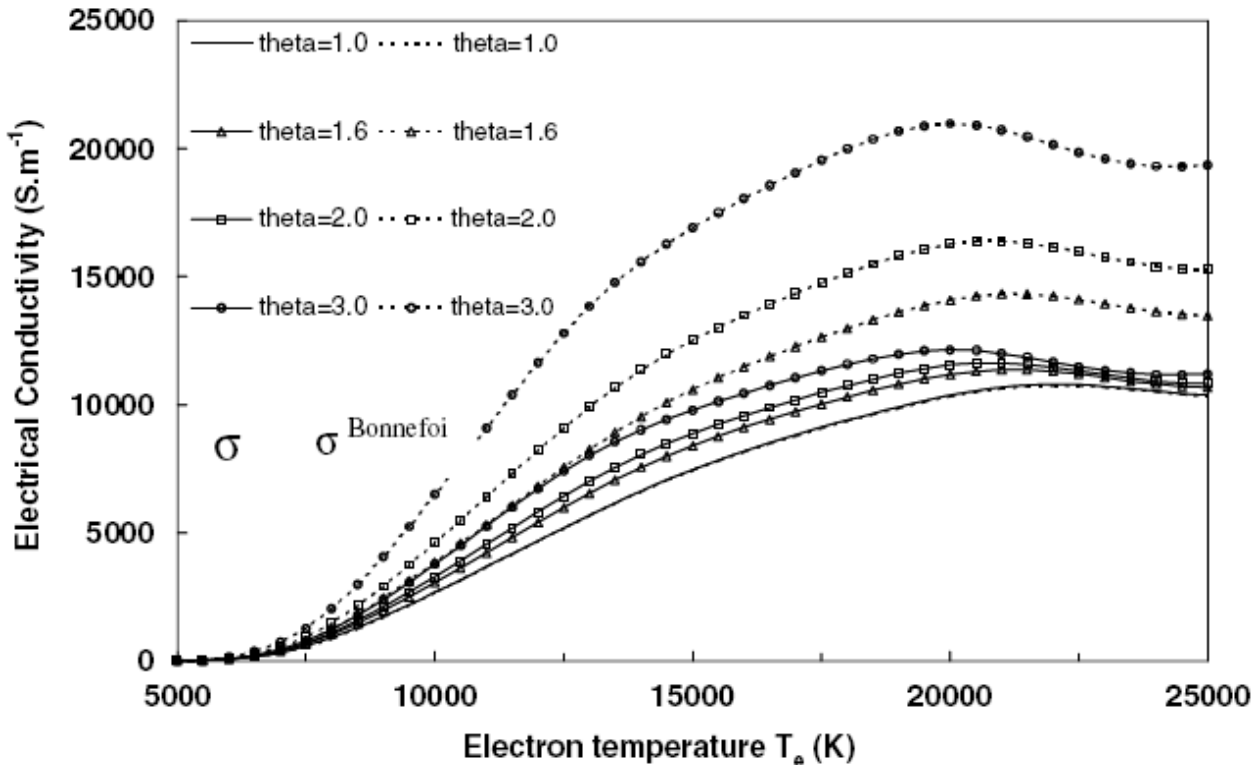

**Fig. 7** Dependence on electron temperature of the electrical conductivity of an atmospheric-pressure argon plasma, calculated for different values of the ratio $T_e/T_h$ using Devoto's approach (Ref 40, 58) modified by Bonnefoi *et al* (Ref 55, 56) and the approach of Rat *et al* (Ref 57) (——)

Devoto (Ref 54) was the first to propose a solution of the Boltzmann equation by decoupling the calculation of transport coefficient of electrons and heavy species. This approximation relies on the large mass difference between the two types of species. Later, Bonnefoi (Ref 55) and Aubreton *et al* (Ref 56) showed that the force vector of diffusion $\mathbf{d}_j$ is not compatible, according to the formulation of Devoto, with the relationship $\sum \mathbf{d}_j = 0$ required for mass conservation, and they introduced a modified formulation. However, more recently, Rat *et al* (Ref 57) showed that conservation of mass is also not guaranteed by these approaches, and the interactions between electrons and heavy species must be considered in the modified Chapman– Enskog method. For example, the introduction of this coupling in the calculation of the electrical conductivity, for the same plasma composition, led to differences of up to 100% or more, as shown in Fig. 7. Figure 7 shows the comparison between the variation with the electron temperature of the electrical conductivity of an atmospheric-pressure argon plasma, for different values of the ratio $\theta = T_e/T_h$, using Devoto's approach (Ref 40, 58) modified by Bonnefoi *et al* (Ref 55, 56) and using the approach of Rat *et al* (Ref 57). At equilibrium both approaches converge, but the difference increases as $\theta$ increases. It is worth noting that Devoto's approach is widely used not only in equilibrium thermal plasma models but also in the modelling of non-equilibrium atmospheric plasma discharges because of the availability of the simplified expressions for transport coefficients as functions of collision integrals, which can be readily implemented within codes. However, Devoto's approach cannot satisfy mass conservation and the simplified expressions provide results (electrical conductivity and translational electron thermal conductivity) that can be quite different from those obtained with a full calculation, i.e. retaining the coupling between electrons and heavy species in the Boltzmann equation. However, the main differences between the different methods of calculation of transport properties arise from the choice of the method of composition calculation as illustrated in Fig. 8.

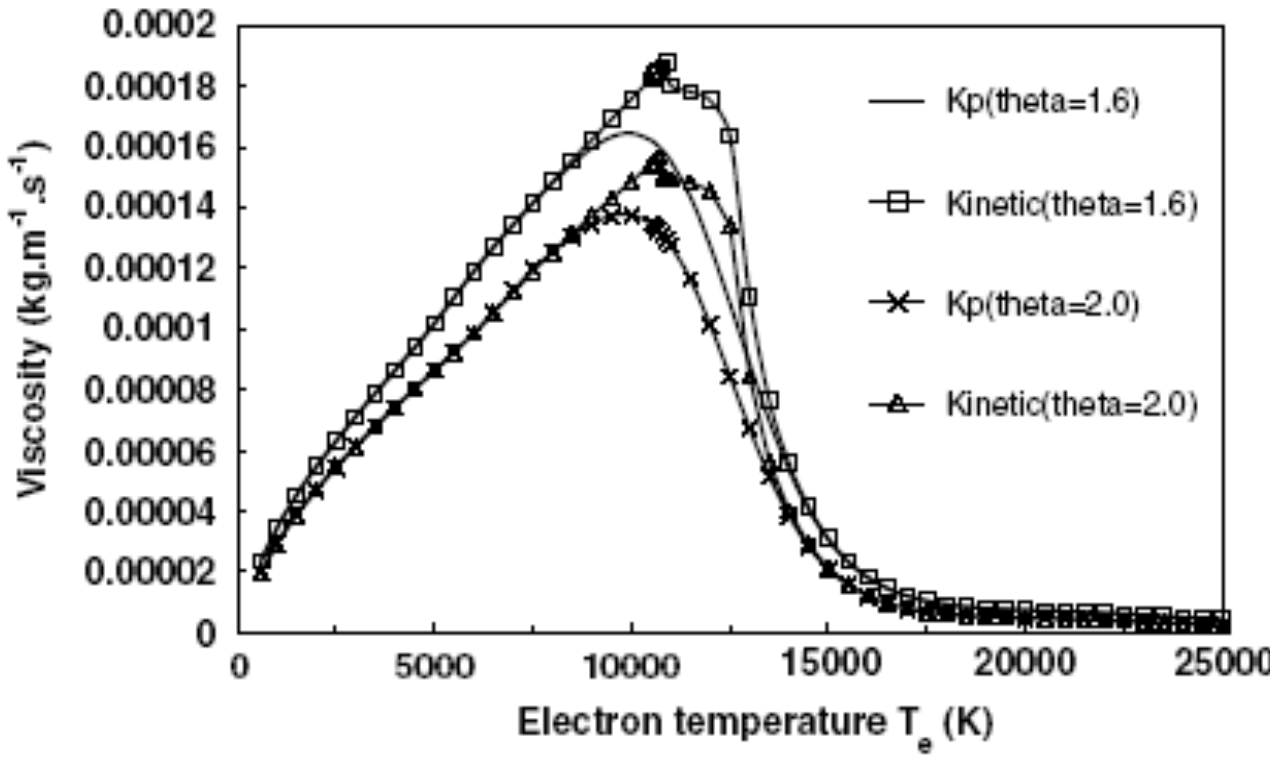

**Fig. 8** Dependence of viscosity on the electron temperature of an Ar–$H_2$ (50% mol) mixture at atmospheric pressure using compositions calculated by the steady-state kinetic calculation (kinetic; section 2.4) and van de Sanden's method for $\theta = 1.6$ and $\theta = 2.0$ (Ref 59)

Figure 8 shows the dependence of viscosity on the electron temperature of an Ar–$H_2$ (50% mol) mixture at atmospheric pressure using compositions calculated by the steady-state kinetic calculation and van de Sanden's method for $\theta = 1.6$ and $\theta = 2$. Due to the delay in ionization introduced by the kinetic method, large discrepancies can be observed between 8000 and 14 000 K; viscosity continues to increase until the ionization regime is reached. The maximum is therefore shifted to higher temperature with respect to van de Sanden *et al*'s method. As in the case of plasma composition, calculations of transport coefficients would also require experimental validation, which is not available yet.

At present, transport properties in NLTE are available for some pure gases (Ar, $O_2$, $N_2$, $H_2$) and their mixtures (Ar–$H_2$, Ar–$O_2$, Ar–$N_2$) with a simplified theory (Ref 41, 60, 61, 56); The application of the theory proposed by Rat *et al* has been already presented for Ar, Ar–He, Ar–Cu and Ar–$H_2$–He plasmas (Ref 38, 62, 63, 64). Air, Oxygen and oxygen-nitrogen transport properties in NLTE have been reported by Gupta et al (65) and Ghorui *et al* (Ref 60, 66) whereas Colombo (Ref 67) performed calculations for $O_2$, $N_2$, and Argon for electron temperature up to 45 000 K.

# 4 Boundary Conditions

## *4.1 Inflow*

Inflow boundary conditions are probably the simplest to implement in a DC arc plasma torch simulation. Inflow conditions are typically specified by imposing values of known properties, typically velocity and temperature, over the region where the gas enters the computational domain (e.g. the left-hand-side region in Fig. 1). Nevertheless, care must be taken when imposing inflow conditions as the type and number of these has to be consistent with the type and number of outflow conditions, as required for the well-posed formulation of compressible flow problems. This implies that the modeler needs to know/assume beforehand the state (subsonic or supersonic) of the inflow(s) and outflow(s). In some cases, the gas near the inflow and outflow regions can be considered incompressible, which simplifies significantly the imposition of boundary conditions in those regions.

The specification of pressure as inflow or outflow condition in arc plasma torch simulations is particularly cumbersome, especially if the simulation domain only covers the interior of the torch. For incompressible internal flows, pressure is often imposed as an outflow condition, whereas specified velocity as inflow condition. For compressible flows, pressure can be used as an inflow or outflow condition depending if the flow is subsonic or supersonic. The inflow in a DC arc torch is often incompressible (very-low Mach number) whereas the flow that leaves the torch is certainly compressible, either subsonic or supersonic. The simulation of DC arc plasma torch flows frequently requires to experiment with different sets of inflow/outflow conditions to find the most appropriate and physically sound conditions.

Inflow boundary conditions often involve the description of the gas injection process. Gas can be injected straight (in the direction parallel to the torch axis), radially, tangentially (i.e. with swirl), and often using a combination of the above. Different forms of gas injection seek to impose different characteristics on the plasma flow, e.g. enhance arc constriction or increase gas mixing. The detailed modeling of the gas injection process is highly desirable, but it is often avoided to reduce the computational cost of the simulation. Gonzalez and co-workers presented in Ref 68 an important analysis of the effects of the accurate versus approximated simulation of the gas injection process in the context of a transferred arc torch simulation.

A difficulty commonly found in thermodynamic non-equilibrium models is the imposition of the inflow condition for the electron energy conservation equation. A simple approach would consist on specifying $T_e = T_h$ at the torch inlet. This approach seems physically correct because the injected gas is certainly in thermodynamic equilibrium. Unfortunately, this approach often produces unrealistic *boundary layers* in the electron temperature field (a severe change in $T_e$ occurs in a very narrow region near the inlet) if the computational domain does not extend sufficiently far upstream of the arc. An alternative approach is to specify a zero gradient condition for the electron temperature:

$$\frac{\partial T_e}{\partial n} = 0 \,, \qquad \text{(Eq 25)}$$

where $n$ represents the normal to the inflow boundary. This approach does not produce boundary layers near the boundary but it does produce unrealistically high electron temperatures if the inflow boundary is too close to the arc. Nevertheless, these unrealistically high electron temperatures do not have physical relevance because the free electron population in the inflow is negligible.

### *4.2 Outflow and Open Boundaries*

The outflow boundary in arc torch simulations is typically the torch exit or some other region downstream the arc, e.g. a region within the extent of the plasma jet. Simulations of the jet produced by non-transferred arc torches, which is characterized by complex dynamics due to the arc movement and large velocity and temperature gradients, require special care in the imposition of outflow conditions.

Ideally, the outflow boundary should be placed far enough from the plasma jet that the remnant velocity and temperature fields are negligible. This approach is seldom followed, especially in three-dimensional simulations due to the increased computational cost of having an extended computational domain. Simulation results from Ref 69 of an axisymmetric steady supersonic jet from a cutting torch are presented in Fig. 9. The results in Fig. 9 clearly show the formation of diamond shocks, the rapid expansion of the jet, and rapid decay of temperature along the jet axis. The computational domain is large enough that no significant flow features leave the domain, and hence, a zero velocity gradients proved adequate as outflow boundary condition.

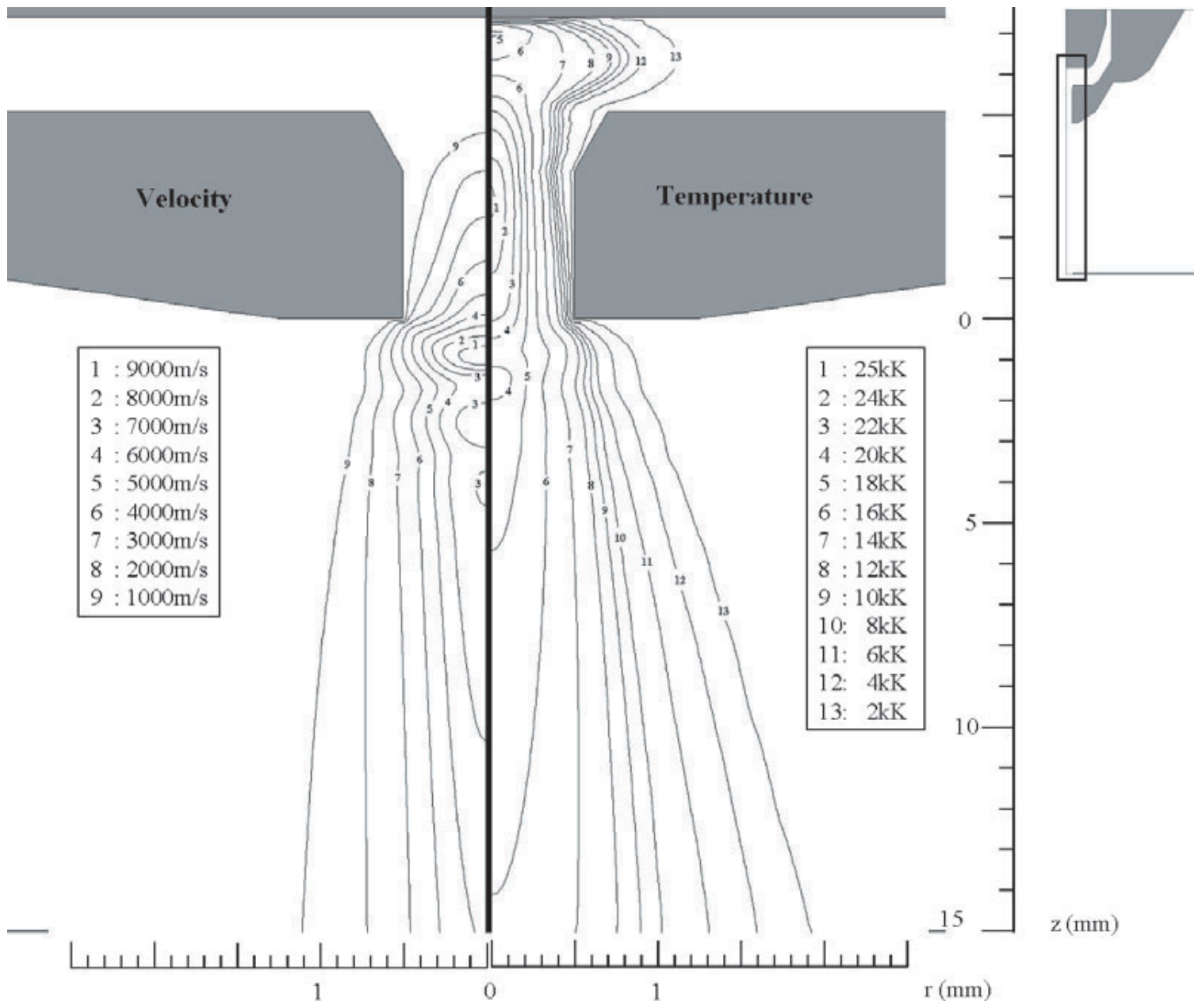

**Fig. 9** Velocity and temperature distribution in a supersonic gas jet issuing from a cutting torch; the diamond shocks can be clearly seen (69)

By far, the most frequently used outflow condition in DC arc plasma torch modeling is the imposition of zero-gradient of the transported variable. This condition is probably the easiest to implement, but unfortunately it is often too reflective, especially when the flow approaching the boundary varies significantly in time and/or space. Other typically used conditions are the imposition of a constant velocity gradient or zero second-order derivative. Outflow conditions need to ensure the uninterrupted transit of the flow characteristics out of the domain. Typical effects of the use of inadequate outflow conditions are unphysical heating, pressure build-up, and wave reflection. The use of physically sound boundary conditions often prevents the first two effects. But, to prevent wave reflection, typically more sophisticated numerical techniques need to be employed.

Many approaches have been developed to avoid the reflection of waves: Non-linear characteristics, grid stretching, fringe methods, windowing, absorbing layers (Ref 70). Among those, the use of absorbing layers (also known as sponge-zones) is probably the simplest and is very often used. The use of an absorbing layer for a given outflow boundary for the transport equation of variable $\psi$ (see Eq 1) consists of modifying the equation near that boundary as:

$$\frac{\partial \psi}{\partial t} + \nabla \cdot \mathbf{f}_{\psi} - s_{\psi} = -\sigma_{\psi} (\psi - \psi_{\infty}), \qquad \text{(Eq 26)}$$

where $\sigma_{\psi} = \sigma_{\psi}(\mathbf{x})$ is a friction coefficient that varies spatially in the direction normal to the boundary, and $\psi_{\infty}$ represents a reference value of $\psi$ (e.g. the value of $\psi$ far from the boundary). The design of $\sigma_{\psi}$ should ensure a smooth transition from 0 in the flow domain to a positive value at the boundary. The region in which $\sigma_{\psi}$ is greater than 0 is known as the *absorbing layer*. A large enough value of $\sigma_{\psi}$ causes disturbances to decay exponentially and at the same time makes the variable $\psi$ approach $\psi_{\infty}$. This method does not completely prevent wave reflection, but it does allow attenuation of outgoing waves and that any reflected wave will continue decaying as it travels through the absorbing layer. The absorbing layer method has successfully been applied to the thermal plasma jet simulations in the work of Trelles *et al* (Ref 71) (see Fig. 10).

Figure 10 shows a time sequence of the dynamics of the arc inside the torch and the plasma jet obtained numerically with a NLTE model, represented by iso-contours of heavy-species temperature, as well as high-speed images of the plasma jet for the same torch and similar operating conditions. The plasma jet presents large-scale structures due to the dynamics of the arc inside the torch (to be explained in Section 5.1), whereas the fine-scale structures are a consequence of the interaction of the jet with the cold surrounding gas. No turbulence model has been employed in those simulation results. The imposition of a sponge zone (i.e. Eq 26) allows the uninterrupted transit of the large and small structures formed by the jet through the boundary.

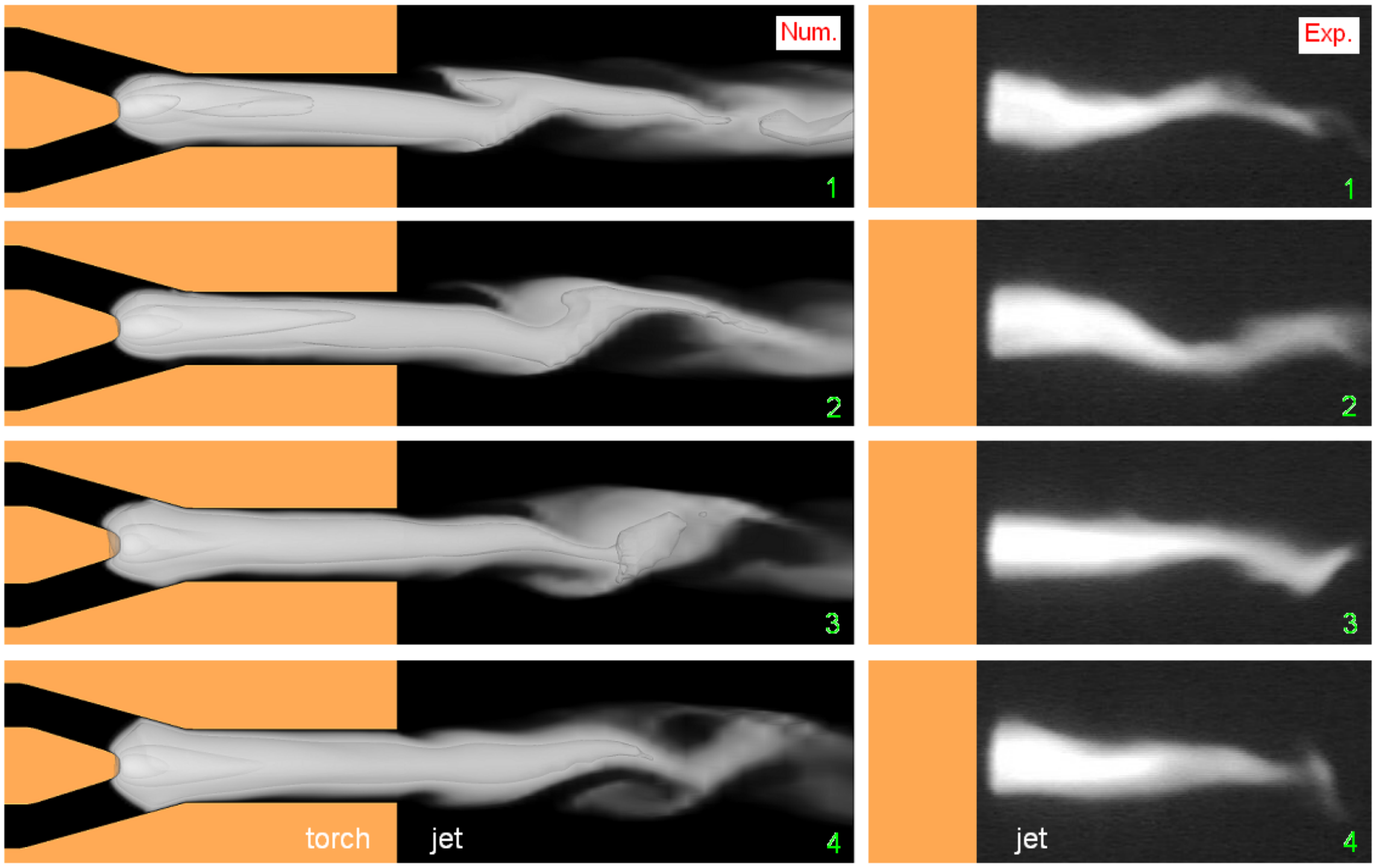

**Fig. 10** Heavy particle temperature distribution obtained with a non-equilibrium model (*left*) and high speed images of the plasma jet (*right*) (Ref 71)

### *4.3 Walls*

The modeling of walls as boundary conditions needs to differentiate between electrodes and non-conducting walls. Non-conducting walls are modeled according to the no-slip condition and the type of heat transfer in place. The modeling of heat transfer from the plasma to conducting boundaries is very involved and will be treated in the next two sections. For non-conducting walls, the major challenge is the modeling of the heat transferred from the plasma. This aspect is particularly crucial in transferred arc torches, where the constricting tube dissipates a large amount of heat from the plasma. In non-transferred arc torches, the regions of the anode far from the arc attachment could also be considered as non-conducting walls as no electric current is conducted through them. The most frequent approach used for the modeling of heat transfer in non-conducting walls of an arc torch system is the use of an overall convective heat transfer coefficient:

$$q_{wall} = -\kappa \frac{\partial T}{\partial n}\bigg|_{wall} = h_w (T - T_w) \, , \quad \text{(Eq 27)}$$

where $q_{wall}$ is the heat transferred to the wall, and $h_w$ and $T_w$ are the heat transfer coefficient and the reference temperature, respectively. In the arc torch modeling literature, $T_w$ is often defined as the cooling water temperature, whereas the coefficient $h_w$ is on the order of $10^5$ W-m$^{-2}$-K$^{-1}$ (Ref 72, 73, 74). For thermodynamic non-equilibrium models, Eq 27 can be used for imposing the heavy-species temperature $T_h$ boundary condition ($T \leftarrow T_h$).

For two-temperature models, it is not evident how to define the boundary condition for non-conducting walls for the electron energy conservation equation. Probably the simplest, and the most often used, approach is to define a *zero electron temperature gradient* condition (e.g. Ref 75).

The specification of electromagnetic boundary conditions is relatively straightforward as it basically consists on specifying the wall as a non-conducting surface, i.e. zero current density.

### 4.4 Anode

The specification of boundary conditions for the anode surface follows the descriptions of the previous section (i.e. non-slip condition), except for the treatment of the energy and electromagnetic boundary conditions.

Plasma flows typically develop what are called plasma sheaths near the electrodes. There are large property variations within these regions that often are negligible within the bulk plasma, like charge accumulation and thermodynamic non-equilibrium. The anode sheath thickness is on the order of a few Debye lengths, where the Debye length is a measure of the charge screening felt by a charged particle due to the other particles (Ref 1, 76). For thermal plasmas, the Debye length is often very small compared to the characteristic length of the flow (e.g. the torch diameter). This causes that the anode influence on the flow be localized very close to the anode surface.

The boundary conditions at the anode surface for the electromagnetic fields often consist of imposing a reference value of electric potential (e.g. $\phi_p = 0$ along the anode surface), whereas the total amount of current transferred is determined by the cathode boundary condition, as explained in the next section. An improved approach consists of including part of the electrodes in the computational domain, and hence solving the energy conservation and electromagnetic equations through the domain conforming the electrodes (e.g. see Ref. 77).

The accurate modeling of anode heat transfer in non-transferred arc torches is very important because erosion due to high heat fluxes often limits the life of the anode. For transferred arc torches the heat transferred to the anode is a measure of the efficiency of the plasma process (e.g. plasma cutting, welding). The description of the heat transfer to the anode in thermal plasma systems is quite involved due to the large number of coupled processes that intervene. For a monatomic gas, with a single type of ion, the total amount of heat to the anode can be expressed by (Ref 78):

$$q_a = -\kappa_h \frac{\partial T_h}{\partial n} - \kappa_e \frac{\partial T_e}{\partial n} + q_e + J_{qe} W_a + J_{qi}(E_i - W_a) + q_r\,, \qquad \text{(Eq 28)}$$

where $q_a$ represents the heat transferred to the anode surface, $J_{qe}$ and $J_{qi}$ are the electron and ion current densities in the direction normal to the anode ($\mathbf{n}\cdot\mathbf{J}_q = J_q = J_{qe} - J_{qi}$), $q_e$ represents the transport of energy by the electron flux, $W_a$ is the work function of the anode material, $E_i$ is the ionization energy of the ion, and $q_r$ is the radiative heat flux to the anode.

The first two terms on the right side of Eq 28 represent the heat transferred by conduction by the heavy species and electrons, respectively. The fourth term represents the electron condensation energy, i.e. the energy transferred to the anode when electrons are incorporated into the lattice of the anode material. The fifth term represents the heat due to ion recombination at the surface; this term is generally small because the ion current is often a small fraction of the total current. Typically, the third and fourth terms and the heavy-species conduction term are the most important ones. Therefore, the total heat to the anode strongly depends on the current density (particularly $J_{qe}$) to the anode. The electron energy transport term is frequently described as:

$$q_e = J_{qe}\left(\frac{h_e}{e} + U_a\right), \qquad \text{(Eq 29)}$$

where $h_e = 2.5k_B T_e$ is the electron enthalpy and $U_\mathrm{a}$ represents the voltage drop across the anode sheath (i.e. the anode fall), which is assumed positive. For the case of negative anode fall, the $U_\mathrm{a}$ term is often dropped from Eq 29. The first term represents the transport of electron energy by mass diffusion (see Eq 7), while the second term describes the electron energy gained in the assumed *free fall* regime in front of the surface. Equations 28 and 29 present the basic components of the modeling of heat transfer to the anode. For NLTE models, the boundary conditions for $T_h$ and $T_e$ can be obtained by splitting Eq 28: The terms involving electrons specify the boundary condition for $T_e$, and similarly for $T_h$. The above description can be extended to include phase change processes of the anode material (i.e. evaporation), and surface reactions; these effects could have a significant consequence on the overall heat transferred.

Figure 11 presents the results of a chemical and thermodynamic non-equilibrium simulation of an argon arc in crossflow performed by Li et al (Ref 79). This flow can be considered as an exemplary flow for the study of the anode attachment inside non-transferred arc plasma torches as it clearly displays the opposing effects of the flow drag and electromagnetic forces. For this flow, the boundary conditions imposed over the anode have a primary effect on the final location and stability of the anode attachment, and therefore can be used to weigh the importance of the different terms in Eq 28.

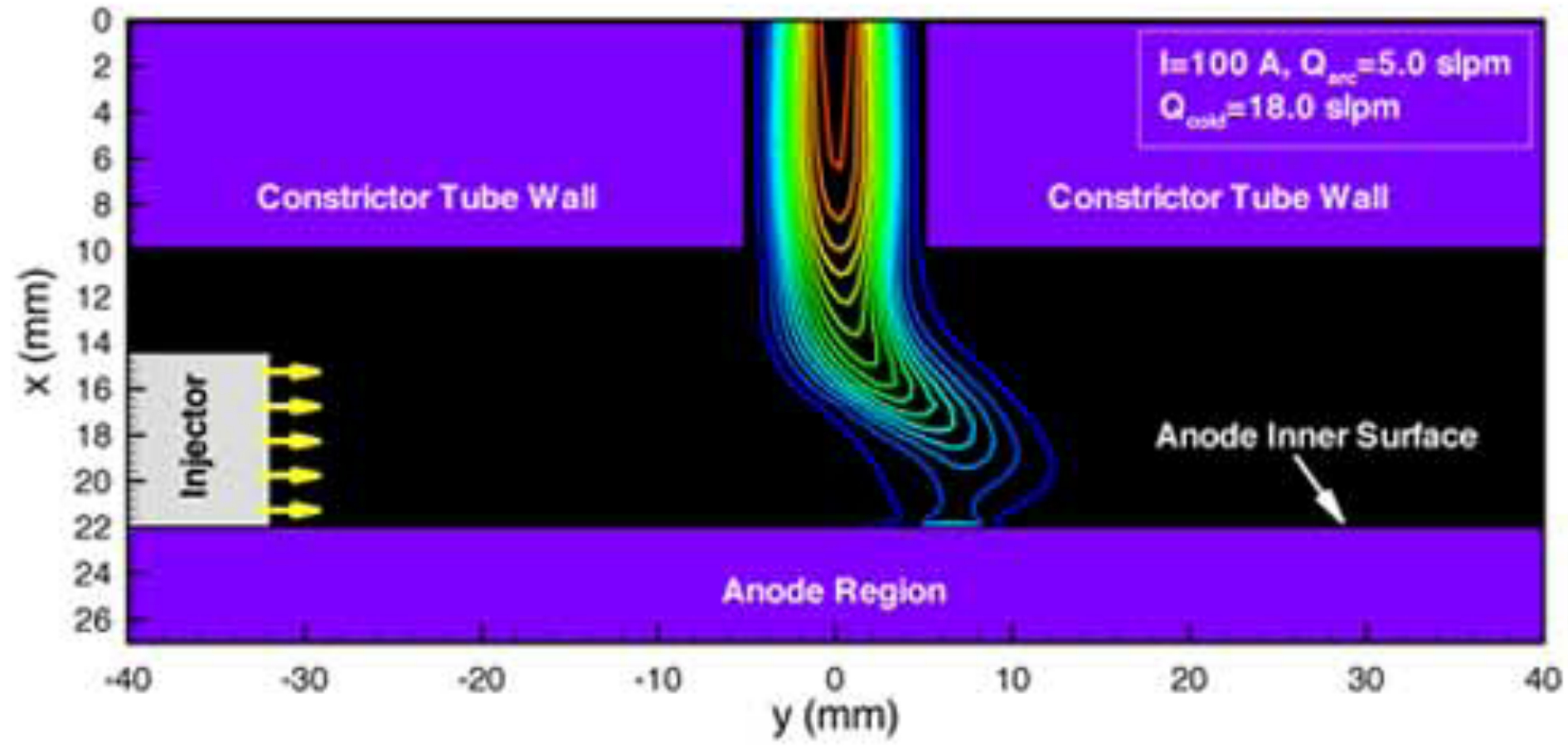

**Fig. 11** Electron number density distribution in arc blown laterally by a stream of cold gas (Ref 79)

When no sheath model is used in a LTE model, due to the thermodynamic equilibrium assumption, the electron temperature is equal to the heavy particle temperature, which is low (i.e. less than 1000 K) near the electrodes due to the intense cooling they experience, especially near the anode surface. Hence the equilibrium electrical conductivity of the plasma, being mostly a function of the electron temperature, is extremely low (i.e. less than 0.01 S/m for most gases), which limits the flow of electrical current through the plasma – electrode interface. To allow current continuity through the plasma-anode interface without the use of a sheath model or a NLTE model, an alternative used in (Ref 74, 80, 81) is to let the temperature remain high enough (i.e. above 7000 K) all the way up to the anode surface. The clear advantage of this approach is that it is consistent with the LTE assumption in the sense that the heavy particle temperature remains equal to the electron temperature (both assumed equal to the equilibrium temperature), which remains high all the way up to the anode surface. Another approach consists of specifying an artificially high electrical conductivity in the region immediately adjacent to the electrodes (Ref 72, 82, 83, 84). The value of this “artificial” electrical conductivity used in the literature is somewhat arbitrary. The only requirement for its value is that it needs to be high enough in order to ensure the flow of electrical current from the plasma to the electrodes. This latter model lets the arc reattach whenever it gets in contact, or "close enough", to the anode surface at an axial location which is more thermodynamically favorable, i.e. a location that produces a configuration of the arc with a lower total voltage drop. However, this type of formation of a new attachment is different from the reattachment process described in Section 5. Important studies of the effect of the anode modeling in an arc plasma flow were performed by Lago *et al* (Ref 23) and later expanded to three-dimensional modeling in Gonzalez *et al* (Ref 85). Their models included the effect of metal vapor on the arc and melting of the anode, detailed heat transfer between plasma and anode (similar to Eq 28), and radiative transfer using view factors.

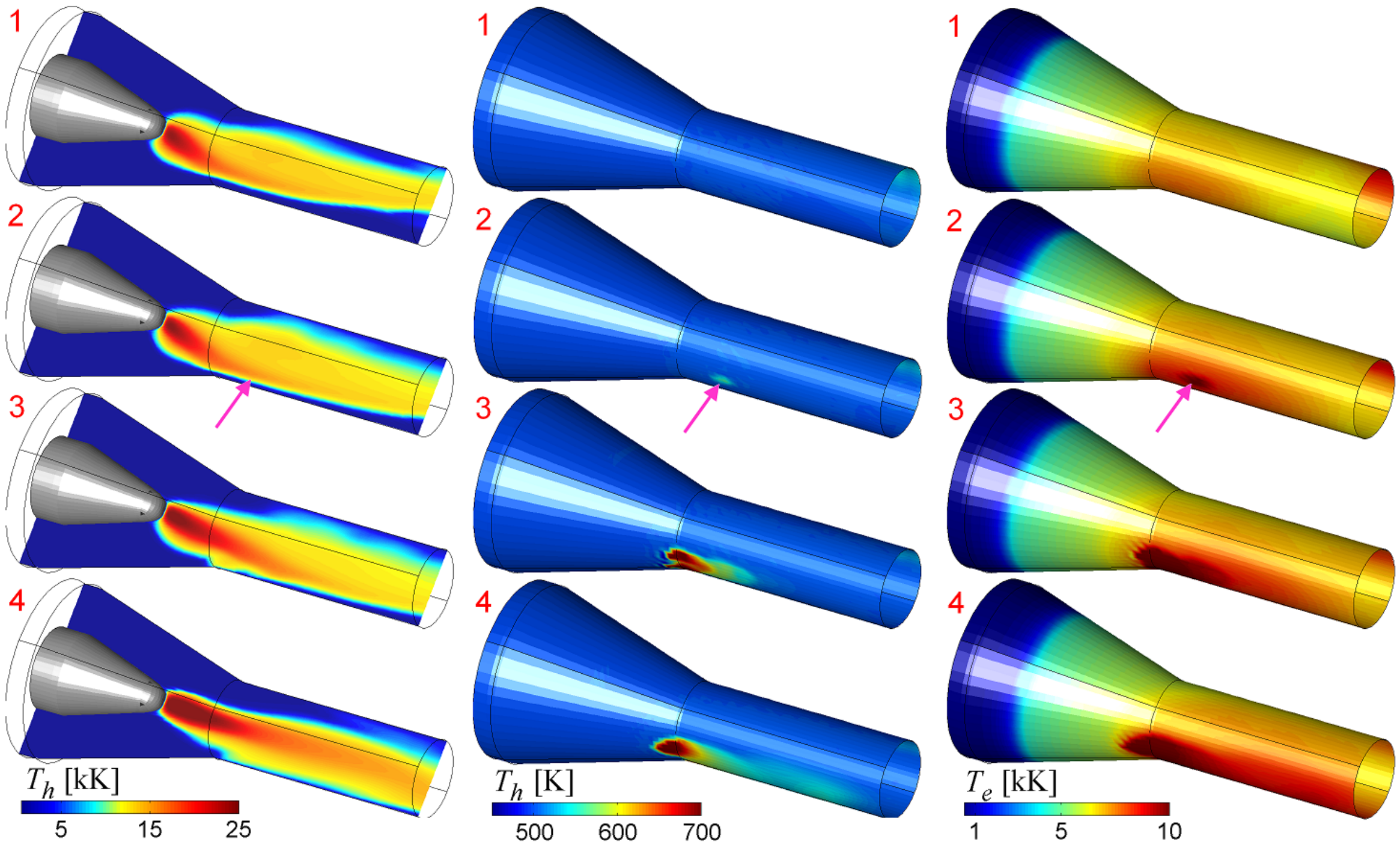

**Fig. 12** Heavy particle temperature across the torch (*left*), and heavy (*center*) and electron (*right*) temperatures close to the anode surface (Ref 87)

Figure 12 shows time-sequences of the distribution of heavy-species temperature inside the torch and heavy-species and electron temperatures ($T_h$ and $T_e$) over the anode surface obtained with a NLTE model. The figure shows the significance of the anode heat transfer depicted by the growth of a new anode attachment spot as well as the large difference in magnitude of the electron and heavy-particle temperatures over the anode. The size of the spot given by the electron temperature distribution is significantly larger than that given by the heavy-particle temperature. The works of Li *et al* (Ref 72) and Park *et al* (Ref 86) clearly display the size and evolution of the anode attachment as obtained by LTE simulations.

### 4.5 Cathode

Cathodes are the source of electrons in thermal plasma torches. The cathode in DC arc plasma torches used for plasma spraying is thermionic; that is the electrons are emitted as a consequence of the high temperature of the cathode. The region in front of the cathode can be divided into two distinctive parts: The ionization region, and the space charge sheath. Similarly to the anode region, these regions are very small compared to the characteristic length of the flow. Typically there is a considerable voltage drop in this thin layer and considerable power is deposited in it. This power is a consequence of the balance between the energy flux of ions and electrons from the plasma to the cathode surface and the heat removed by the electrons leaving the cathode (Ref 88).

The accurate modeling of the cathode region is quite involved due to the variety of chemical and electrical phenomena taking place. Furthermore, it has been shown that evaporation of the cathode material can have a significant effect on the plasma flow dynamics (Ref 89). Indeed, the metal vapor increases significantly the electrical conductivity of the plasma in front of the cathode, which causes constriction of the arc. These effects have to be added to the stability of the cathode spot (the region with highest current density), which is often of primary importance in cathodes whose geometry does not favor a preferred spot (see Ref 90 for detailed modeling of cathode spot stability).

The large computational cost associated to the self-consistent modeling of the electrode regions and the plasma flow has prevented them to be widely used in arc plasma torch simulation. A distinctive example of the coupled modeling of electrodes and thermal plasma flow in an industrial application is the work of Khokan *et al* (Ref 91) of the simulation of the arc discharge in a HID lamp. In their model the cathode region is modeled using the nonlinear surface heating model of Benilov (Ref 88). The work by Li and Benilov (Ref 92) of the coupled simulation of the arc and cathode region revealed that the electric power deposited into the cathode region is transported not only to the cathode, but also to the arc column.

The need to reduce the computational cost of the modeling of the electrode regions in industrial thermal plasma flows has motivated the development of different *sheath* models. These models try to describe in a simplified manner the main physical effects that dominate the electrode – plasma interface. Particularly significant is the unified approach developed by Lowke *et al* (Ref 93) which, when applied to the modeling of the cathode, does not require the specification of a current density profile.

Probably the most common (as well as least expensive) approach used to model the cathode boundary is to specify the current density profile as boundary condition for the electromagnetic equations. A common profile used in the DC arc plasma modeling literature has the form:

$$J_{cath} = J_{cath0} \exp(-(\frac{r}{R_{cath}})^{n_{cath}}) \qquad \text{(Eq 30)}$$

where $J_{cath}$ is the current density over the cathode surface, $r$ the radial coordinate measured from the torch axis, and $J_{catho0}$, $R_{cath}$, and $n_{cath}$ are parameters that control the shape of the profile, preferably to mimic experimental measurements. These parameters are not independent as the integration of the current density profile over the cathode should be equal to the total current imposed. Typically, for commercial plasma spray torches operating between 100 and 800 A, $J_{catho0}$ is of the order of $10^8$ A/m$^2$, the exponent $n_{cath}$ varies between 1 for acute conical cathodes (Ref 68, 72) to ~ 4 for more rounded ones (Ref 84), and the characteristic distance $R_{cath}$ is typically less than 1 mm.

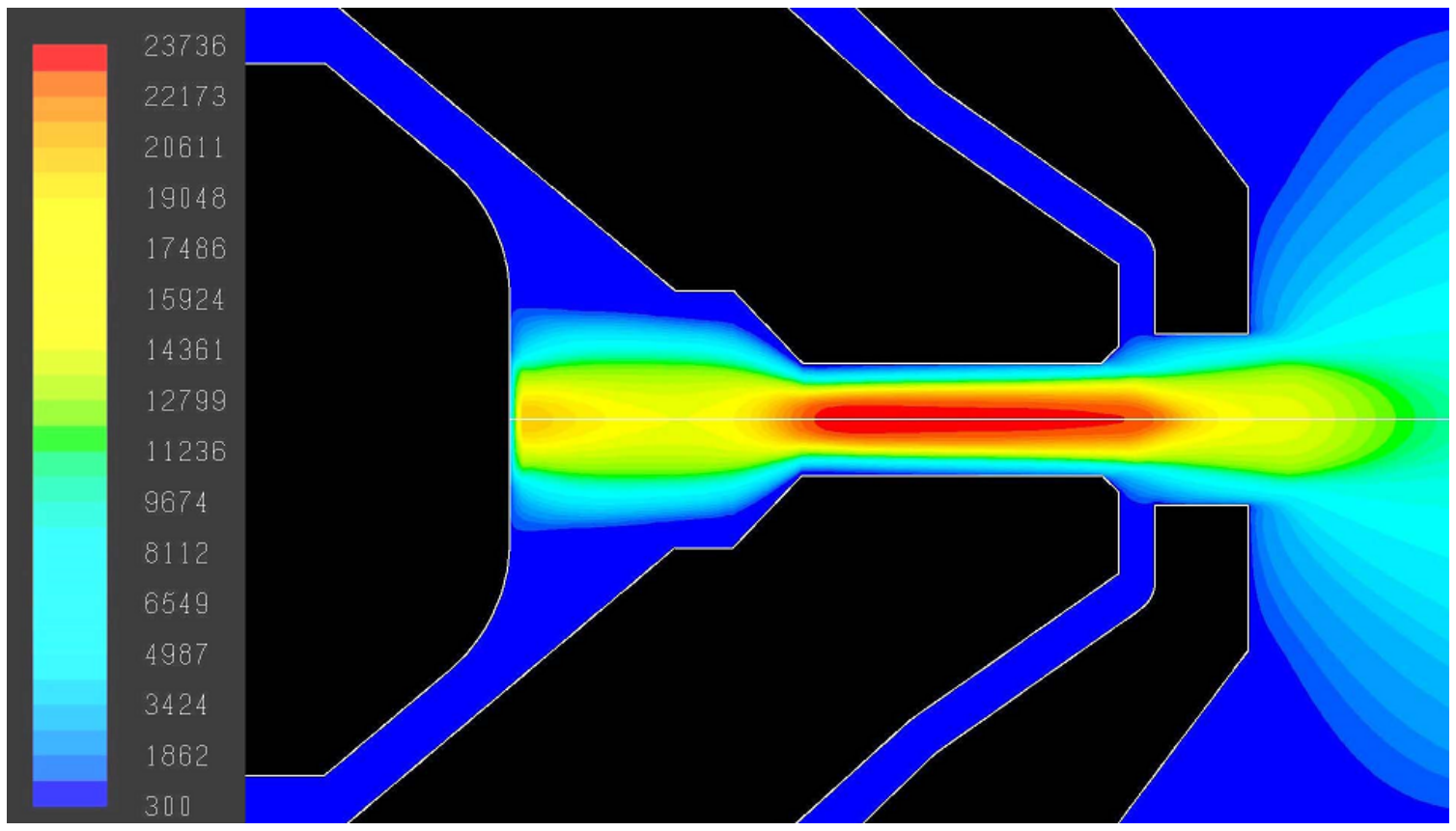

**Fig. 13** Comparison of plasma temperature [K] for a commercial plasma cutting torch with and without magnetization apparatus; total current = 120 A, nozzle diameter 1.37 mm (Ref 94)

As boundary conditions for the energy conservation, often a specified equilibrium or heavy-species temperature distribution is imposed over the cathode surface (where the highest temperature is usually assumed close to the melting point of the cathode material (e.g. ~ 3600 ºC for tungsten), whereas a zero-gradient condition is imposed for the electron temperature. These conditions are rough approximations, and when possible, an adequate cathode region model should be employed.

The modeling of the cathode region is particularly important in simulations of plasma cutting torches due to the relatively large cathode tip area, relatively small flow volume, and large current densities over the cathode. Fig 13 presents simulation results by Colombo *et al* (Ref 94) of the flow through a commercial plasma cutting torch. The pronounced constriction of the arc, crucial for attaining a precise cut, can be observed as well as a localized high temperature region in front of the cathode.

# 5. Arc Reattachment Models

## *5.1 Operating Modes in Non-transferred Arc Torches*

The dynamics of the arc inside the torch are mostly the result of the imbalance between the electromagnetic (or Lorentz) forces, produced by the local curvature of the current path and the self-induced magnetic field, and the flow drag, caused by the interaction of the incoming cold gas and the hot, low-density, arc. Because the total voltage drop across the torch is approximately linearly-dependent on the arc length, the variation of the total voltage drop over time gives an indication of the arc dynamics inside the torch. The characteristic features of the voltage drop signal over time for given operating conditions have led to the identification of three distinct modes of operation of the torch (Ref 95, 96, 97, 98), namely:

- *Steady*: Characterized by negligible voltage fluctuations and, correspondingly, an almost fixed position of the arc attachment. This mode is not desirable due to the rapid erosion of the anode.
- *Takeover*: Characterized by (quasi-) periodic fluctuations of voltage drop and a corresponding (quasi-) periodic movement of the arc. The spectrum of the voltage signal presents several dominant frequencies. This operating mode is the most desirable because it allows an adequate distribution of the heat load to the anode, and produces well defined arc fluctuations.
- *Restrike mode*: Characterized by a highly unstable, relatively unpredictable movement of the arc and quasi-chaotic, large amplitude, voltage fluctuations. An arc operating in this mode is very unstable and relatively unpredictable; the arc reattachment phenomenon plays a dominant role in the arc dynamics.

For a given torch, the flow can change from the steady mode, to the takeover, and then to the restrike mode as the mass flow rate is increased or as the total current is decreased; or more precisely as the enthalpy number $N_h$ increases (i.e. $N_h$ is proportional to the mass flow rate and inversely proportional to the total current squared). Therefore, the operating conditions determine the dynamics of the arc inside the torch.

### 5.2 Arc Reattachment Process

Even though the exact mechanisms driving the reattachment process are not completely known yet, the relatively high electric fields, added to the abundance of excited species around the arc (i.e. due to UV excitation), and the short time scale of the process, suggests that the arc reattachment is initiated by a streamer-like breakdown. In a streamer-like breakdown, the new attachment starts with a continuous avalanche of UV-excited electrons. This streamer connects the arc with the anode in a region somewhere upstream of the existing arc attachment, creating a conducting channel that allows the establishment of a new attachment. As the new configuration of the arc has a lower voltage drop, it is thermodynamically more favorable, and therefore the new attachment remains over the old one, which is dragged away by the flow. Experimental images of the reattachment process, together with a schematic representation of the phenomena involved, are shown in Fig. 14. Figure 14 shows the displacement of the anode attachment, the formation of a new attachment (i.e. a reattachment process), which causes the momentarily splitting of the current path, and the predominance and further movement of the new attachment.

To mimic the physical reattachment process, a model mainly faces the questions of *where* and *how* to introduce the new attachment. *Where* to locate the new attachment translates into the definition of an adequate *breakdown condition*, whereas *how* to introduce the reattachment translates into the definition of *adequate modifications of the flow field* in order to mimic the formation of an attachment.

The detailed modeling of this process is unfeasible with actual computational methods and computer power, especially when the reattachment process is part of the modeling of a realistic plasma application. The work by Montijn *et al* (Ref 99) is a notable example of the simulation of streamer propagation. A similar model would have to be integrated to an arc flow simulation to realistically simulate the arc reattachment phenomenon. Therefore, for the simulation of industrial thermal plasma applications, approximate models are needed which imitate the effects of the reattachment process within the flow field. Sections 5.3 and 5.4 describe two approaches that have been successfully applied to the simulation of commercial plasma spray torches.

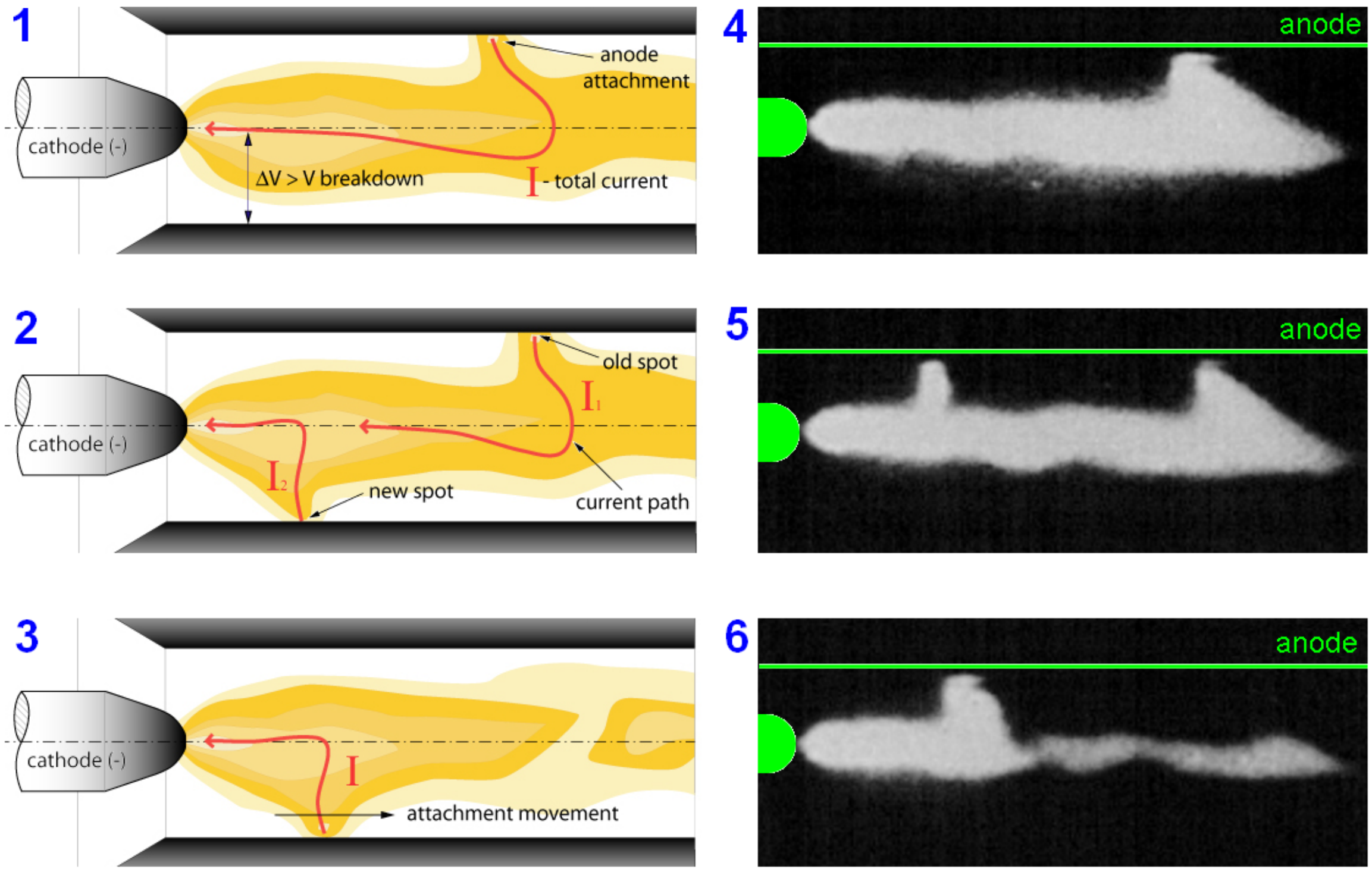

**Fig. 14** Schematic representation of the arc reattachment process (*left*) and experimental high speed images (*right*) from (Ref 95)

### *5.3 Conducting Channel Reattachment Model*

This model, developed in (Ref 84), relies on the fact that the dragging of the arc by the flow field causes the local electric field around the arc to increase. When the local electric field $E_n$ in the direction normal to the anode surface exceeds some pre-specified value $E_b$ (model parameter), namely:

$$\mathbf{E}\cdot\mathbf{n}_a\big|_{\max} = E_{n,max} > E_b \,, \qquad \text{(Eq 31)}$$

where $\mathbf{n}_a$ is the normal to the anode surface, the breakdown condition is met. Centered in that location and in the direction normal to the anode surface, a cylindrical region connecting the arc and the anode is specified. Within that cylindrical region, the electrical conductivity is modified according to:

$$\sigma \leftarrow max(\sigma, \sigma_b)\,, \qquad \text{(Eq 32)}$$

where $\sigma_b$ is an artificially high electrical conductivity and $\sigma$ in the right hand side represents the local electrical conductivity of the plasma. The reattachment process is mainly driven by the specification of $E_b$; the specific value of $\sigma_b$, as well as its spatial variation ($\sigma_b = \sigma_b(\mathbf{x})$) only affects the speed of the process (e.g. larger values of $\sigma_b$ produce a faster reattachment). Moreover, if the physical reattachment process is indeed triggered by the value of the local electric field exceeding certain *breakdown* voltage ($E_b$ in the model), it is reasonable to expect that the value of this breakdown voltage will be a function of the gas composition, and probably only a weak function of the torch characteristics and operating conditions (i.e. nozzle diameter, mass flow rate, total current).

Because of the *free* parameter $E_b$, this reattachment model cannot predict the operating mode of the torch. The model can only predict the arc dynamics inside a torch operating under given operating conditions *and* a given value of $E_b$. If the model is used to simulate a torch operating under conditions leading to a takeover (or steady) mode, a high enough value of $E_b$ should be used in order to ensure that a restrike-like reattachment does not occur. Otherwise, the voltage signal obtained could resemble that of the restrike mode.

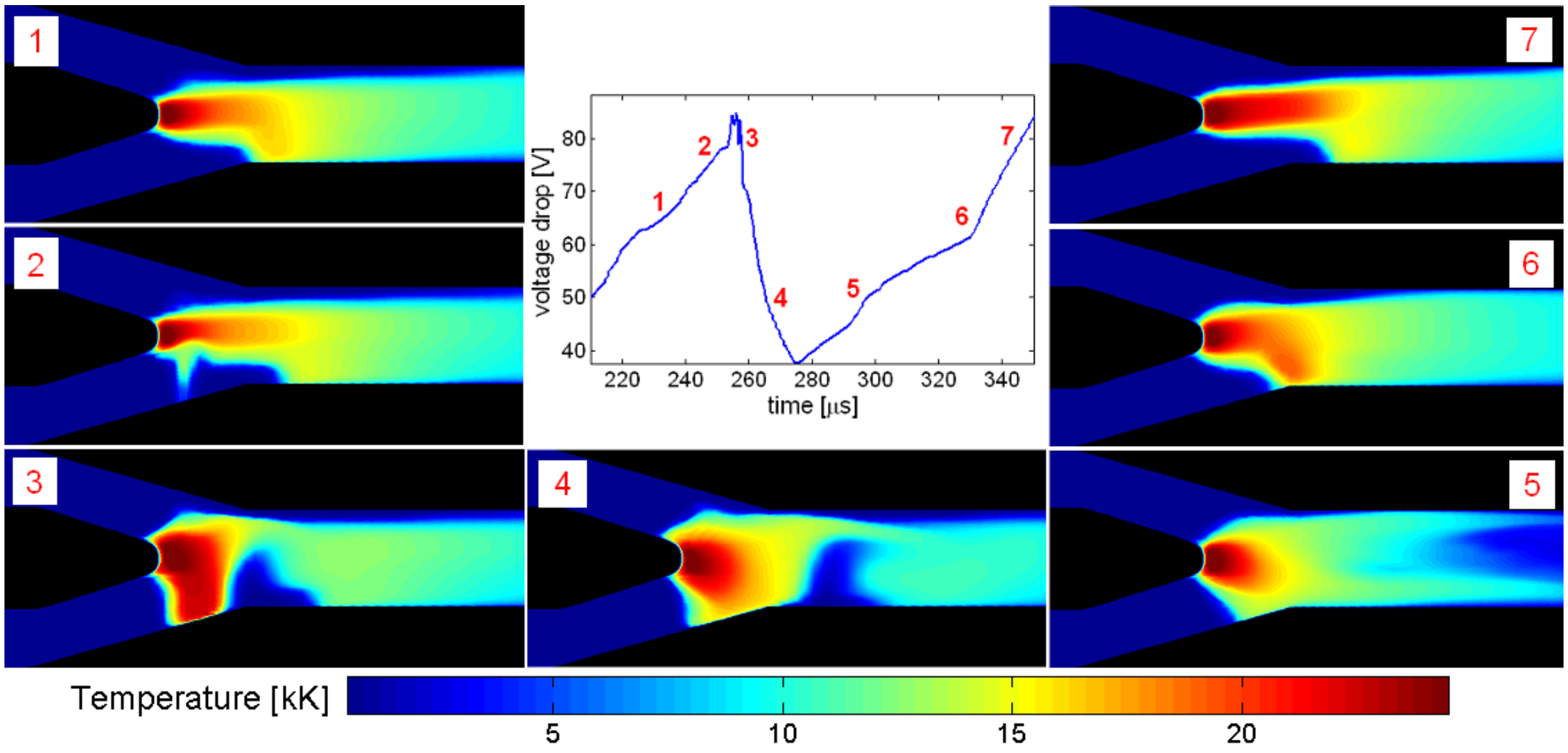

**Fig. 15** Temperature distribution and voltage drop during the reattachment process obtained by a LTE model complemented with a Trelles et al reattachment model (Ref 84)

The time evolution of the temperature distribution and total voltage drop during the formation of a new arc attachment in a commercial torch operating with Ar-He, 60 slpm, 800 A, and straight gas injection are shown in Fig. 15 for $E_b = 5 \cdot 10^4$ V/m. When the local electric field is below the pre-specified value of $E_b$, the arc is dragged by the flow, and its length and the total voltage drop increases linearly (from frame 1 to 2). As soon as the breakdown condition is met (Eq 31), the reattachment model is applied. Frame 2 shows the growing of a high temperature appendage in the direction towards the anode, mimicking the formation of a new attachment. In frame 3, the new attachment forms and, as indicated by the voltage drop between 3 and 4, rapidly overcomes the old attachment, which is dragged away by the flow. Frames 5 to 7 show the dragging of the new attachment accompanied by an approximately linear increase of the total voltage drop. The rapid decrease of voltage drop from 3 to 4 resembles the restrike mode. The results reported in Ref 84 suggest that, by tuning the value of $E_b$, the principal parameter in this reattachment model, one could be able to match the voltage signal obtained experimentally.

## 5.4 Hot Gas Column Reattachment Model

The reattachment model developed by Chazelas *et al* (Ref 80) is also based on the fact that the stretching of the arc column due to the drag forces exerted by the plasma forming gas flow leads to an increase of the voltage drop between the arc column fringes and the anode surface. The model considers that the boundary layer surrounding the arc column breaks when the voltage, and so the electric field, overcomes a certain threshold value. The breakdown process is modelled as follows:

(i) The thickness $\delta$ of the boundary layer that covers the anode surface is defined by the thickness between the region of the flow with an electrical conductivity lower than 150 S/m and the anode surface. For Ar-$H_2$ plasma-forming gas, the value of 150 S/m corresponds to a critical temperature $T_c$ (see Section 3.1, Fig. 4) under which the plasma gas acts as an insulating layer.

(ii) The electric field between the edge of the arc column and the anode wall $E_\ell = (\phi_0 - V_a)/\delta$ (where $\phi_0$ is the potential in the plasma and $V_a$ the potential at the anode surface) is calculated in the whole boundary layer and compared with a critical field $E_b$, under which no breakdown process can occur. A value of $E_b$ of around few $10^4$ V/m had been chosen because it is now well established that the critical electric field is decreased by one order of magnitude when temperatures higher than 3000 K are encountered.

(iii) When the value $E_b$ is reached at a particular location $M$, a short-circuit occurs and a new arc attachment at the nozzle wall appears.

A simple model is used for the ignition of a new arc root attachment by re-arcing. It consists of imposing, a *high-enough* gas column temperature that connects the arc column fringe to the anode wall, at the location where the electric field was found to be greater than $E_b$.

The time-evolution of the temperature distribution and total voltage drop during the formation of a new arc attachment in a commercial torch operating with Ar-$H_2$, 45-15 slpm, 600 A, 6-mm nozzle diameter and straight gas injection are shown in Fig. 16 for $E_b = 5 \cdot 10^4$ V/m. When the local electric field is below the pre-specified value of $E_b$, the arc is dragged by the flow, and its length and the total voltage drop increase linearly (from frame 1 to 2). As soon as the breakdown condition is satisfied, the reattachment model is applied (frames 2 and 3). Frame 3 shows the growing of a high temperature appendage in the direction towards the anode, mimicking the formation of a new attachment. In frame 3, the new attachment is formed and, as indicated by the voltage drop, rapidly overcomes the old attachment, which is dragged away by the flow. Frame 5 shows the dragging of the new attachment accompanied by an approximately linear increase of the total voltage drop.

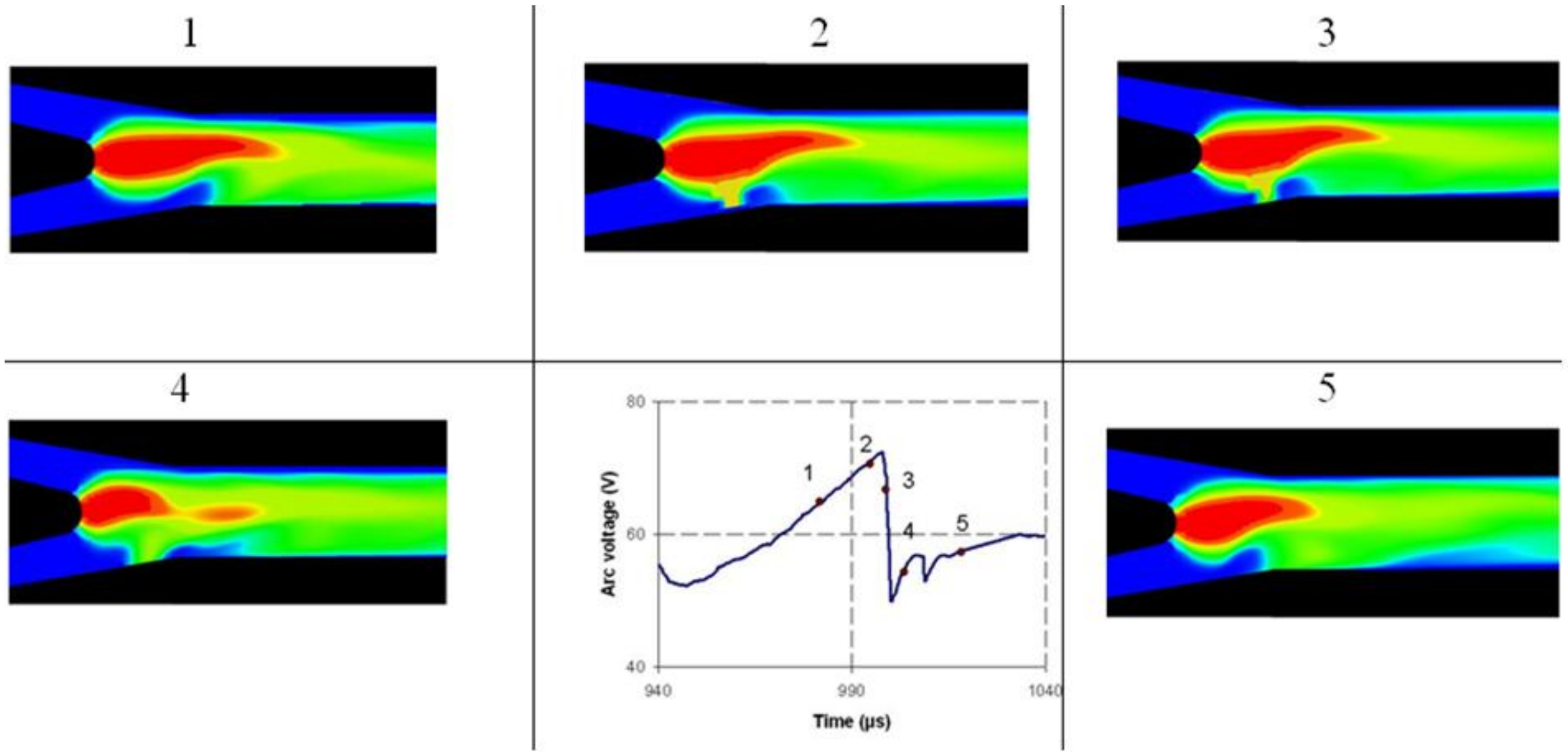

**Fig. 16** Temperature distribution and arc voltage during the reattachment process obtained by a LTE model complemented with a Chazelas' et al reattachment model (Ref 80)

Each breakdown is associated with a negative jump of the voltage. In the case shown in Fig. 16, the latter was found to be 20 ± 5 V. The peak occurs at intervals of about 80 µs, i.e. at an average frequency of about 13 kHz. The predicted time-average torch voltage is about 65 V, close to the actual one of 60 V experimentally measured. The results reveal that decreasing the arc current intensity or increasing the plasma gas flow rate results in an increase of the average boundary layer thickness $\delta$, favouring higher voltage jump amplitude $\Delta\phi$ (Ref 100, 81). As mentioned by Trelles *et al*, (Ref 84) tuning the value of $E_b$, the principal parameter in the reattachment model, permits to either match the frequency or the voltage jump amplitude, the other quantity moving in the opposite manner.

### *5.5 Reattachment in Non-Equilibrium Models*

To the best knowledge of the authors, no reattachment model has been coupled to a non-equilibrium plasma torch simulation yet. This may be due in part to the complexity and computational cost of non-equilibrium simulations and the added difficulty to attain convergence when a reattachment model is used. Figure 17 presents snapshots of simulation results by Trelles *et al* (Ref 75) of the reattachment process obtained with an LTE and a NLTE model for a torch operating with argon, 60 slpm, and 400 A, and swirl injection. The reattachment model of Ref 84 is used in the LTE model, whereas no reattachment model is used in the NLTE model. The figure shows how the anode attachment moves upstream initially; then, due to the net angular momentum over the arc, the curvature of the arc increases, pushing the arc towards the opposite side of the original attachment. These dynamics are interrupted by the occurrence of the arc reattachment process.

Interestingly, the NLTE simulations (which use no reattachment model) display the growth of a high temperature appendage (see arrows in the $T_h$ and $T_e$ plots in Fig. 17) in a region upstream after the point of maximum total voltage drop (which could be correlated with the maximum value of electric field). The formation of the high temperature appendage seems to be driven by high values of the local effective electric field and high values of electron temperature. Even though swirl injection is used, due to the relatively short time scale of the reattachment process, the arc reattaches at almost the opposite side of the original attachment (Ref 101). The reattachment process occurs in a natural manner mimicking the steady and/or takeover modes of operation of the torch. It must be noted that arcs in pure argon as simulated here rarely display a restrike behavior because the boundary layer is usually rather thin. It is expected that non-equilibrium simulations of the restrike mode will require the use of a reattachment model to produce more accurate results.

As explained previously (Section 4.4), in a LTE model a reattachment can occur either due to the application of the reattachment model (i.e. when the breakdown condition is satisfied) or due to the arc dynamics causing the arc to get "close enough" to the anode. The growth of a high temperature region from the arc column towards the anode can be observed in the LTE results in Fig 17 which eventually initiates the formation of a new attachment. Moreover, the application of the reattachment model clearly disrupts the flow significantly.

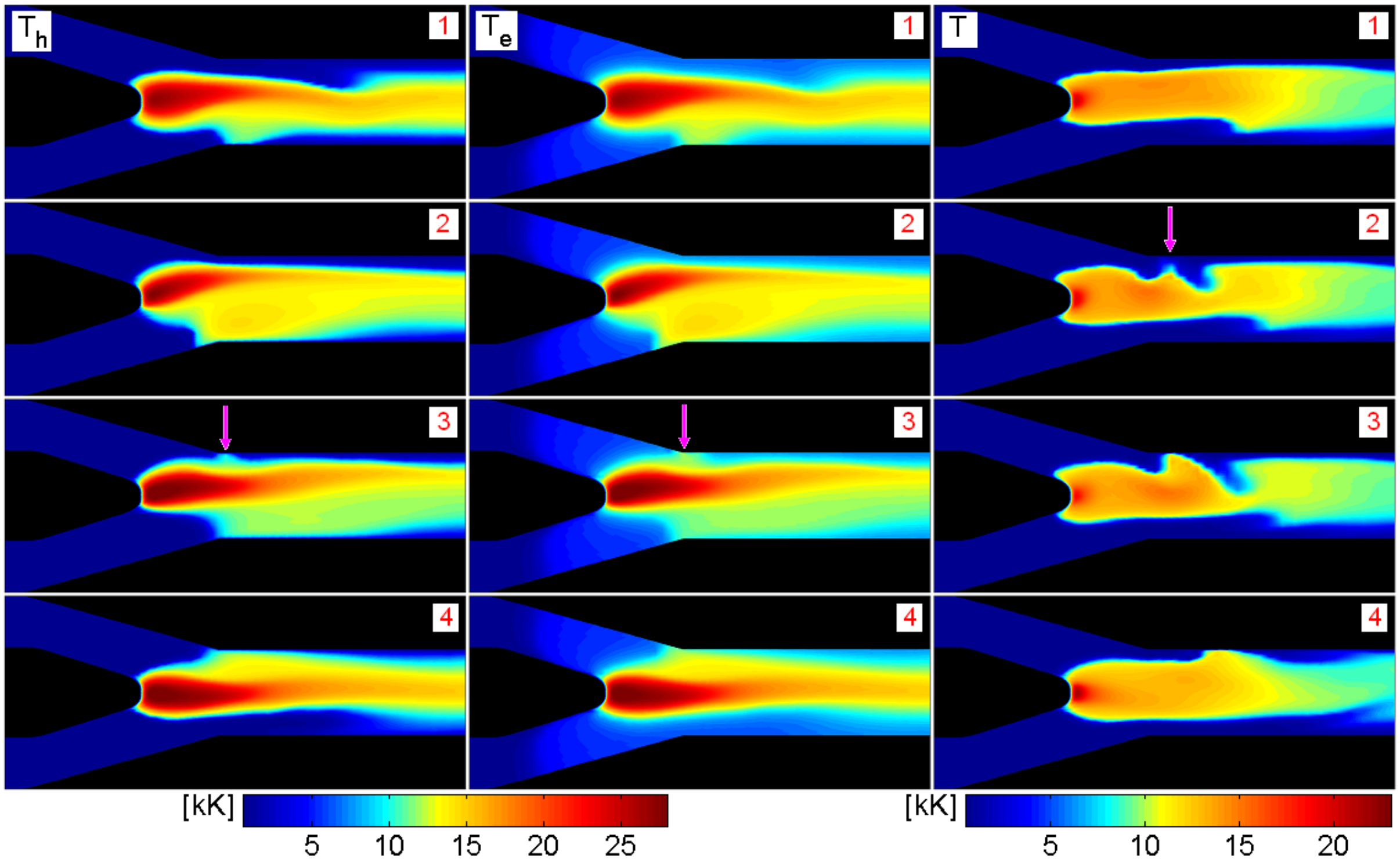

**Fig. 17** Heavy particle temperature (*left*), electron temperature (*centre*) obtained with a non-equilibrium model, and equilibrium temperature (*right*) obtained with an equilibrium model during the reattachment process (Ref 75)

## 6. Conclusions and What's Next

Great progress has been achieved in the simulation of DC arc torches. We have reached a state in which three-dimensional and time-dependent simulations with detailed geometry description of industrial torches are reaching widespread use. These simulations have helped to achieve a better understanding of the operation of DC arc torches and have sometimes led to improved torch designs and plasma processes. However, these simulations still lack of complete predictive power, especially for the simulation of non-transferred arc torches. Several improvements could be achieved with higher computing power and massive use of parallel computing. But most importantly, the complexity of the models needed to describe the different processes that take place, has been the limiting factor for more detailed and accurate plasma torch simulations. The ready availability of different "physics modules" in commercial CFD software has eased the incorporation of several physical processes and a wider range of models into plasma torch modeling. Unfortunately, the models in commercial CFD software are most often not implemented with thermal plasma flow applications in mind, and therefore, often rely on unrealistic assumptions (e.g. turbulence models that assume constant thermodynamic and transport properties, neglect of electromagnetic forces, etc.). Nevertheless, users of commercial CFD software frequently need to develop user-defined routines to be integrated into their simulations in order to account for the missing physical models. In this regard, the use of research, in-house, developed software generally provide the most faithful models for plasma torch simulation (e.g. electrode boundary models, electromagnetic equations). To the previous exposition, we have to add that several processes involved in the description of the plasma flow are not yet understood at the level that accurate models are available. A paramount example of this is the initiation of arc reattachment phenomena.

Some of the developments we anticipate will greatly improve the predictive power of DC arc plasma torch simulations are:

- Widespread use of thermodynamic and chemical non-equilibrium models. These non-equilibrium models necessarily need to use sound values of non-equilibrium thermodynamic and transport properties. The accurate modeling of these properties still represents a big challenge, both in terms of implementation and on computational cost.
- Incorporation of the electrodes into the computational domain. This would have a significant effect in the boundary conditions for electromagnetic equations.
- Detailed modeling of electrodes and electrode processes, particularly heat transfer mechanisms and electric current flows. Furthermore, surface chemistry and phase change phenomena (e.g. electrode material evaporation, anode erosion) should be incorporated into these models.
- Use of more faithful geometry representations. This is particularly important for the analysis of commercial plasma torches and for their design optimization.
- In the case of non-transferred arc torches, more physical and mathematically sound models of the arc reattachment process are needed. It is reasonable to expect that the incorporation of such models into a thermodynamic non-equilibrium plasma torch simulation will be able to reproduce the steady, takeover, and restrike modes of operation without the need for tuning parameters.
- Regarding the modeling of turbulence, DNSs would be highly desirable, especially in order to understand the mechanisms for turbulence formation inside the torch, particularly the role of fluid-dynamic, thermal, and electromagnetic instabilities and the arc reattachment process. DNS data would also guide the development of turbulence models (LES, RANS, and DES) suitable for thermal plasma flow simulations. Furthermore, detailed turbulence modeling would be of great benefit for the understanding of plasma – powder interaction, especially for ultra-fine and nano-scale powders, as these processes are influenced by the fine scale structures of the flow. Additionally, detailed comparison of simulation results against experimental measuremens of turbulent flow characteristics (e.g. correlations, mean quantities, dissipation rates) is required to validate any turbulent thermal plasma flow model.

- Rigorous validation with experimental data. The recent availability of high-fidelity three-dimensional and often time-dependent experimental data, such as the analysis of the anode attachment region in Ref 102, or of the plasma jet in Ref 103, raises the quality, validity, and resolution expected from numerical simulations.

We expect that direct current arc plasma torch modeling will be playing an increasingly important role in the design of thermal plasma processes. Several industrial applications will obtain better yields, higher efficiencies, and improved economical advantage thanks to the systematic use of numerical simulations to guide and/or aid the design and optimization of their processes.